\documentclass[prc,aps,floatfix,twocolumn,showpacs,nofootinbib,superscriptaddress]{revtex4}

\usepackage{amsmath}
\usepackage{amssymb}
\usepackage{amsfonts}
\usepackage{graphicx}
\usepackage{color}

\allowdisplaybreaks     

\renewcommand{\vec}[1]{{\mathbf{#1}}}
\newcommand{\nuc}[2]{$^{#1}${#2}}

\newcommand{\vnabla}{\boldsymbol{\mathbf\nabla}}
\newcommand{\vsigma}{\boldsymbol{\mathbf\sigma}}
\renewcommand{\tensor}{t}

\newcommand{\etal}{\emph{et al.}}
\newcommand{\nn}{\nonumber}

\begin{document}

\title{Tensor part of the Skyrme energy density functional.\\
       II: Deformation properties of magic and semi-magic nuclei}

\author{M. Bender}
\affiliation{Universit{\'e} Bordeaux,
             Centre d'Etudes Nucl{\'e}aires de Bordeaux Gradignan, UMR5797,
             F-33175 Gradignan, France}
\affiliation{CNRS/IN2P3,
             Centre d'Etudes Nucl{\'e}aires de Bordeaux Gradignan, UMR5797,
             F-33175 Gradignan, France}

\author{K. Bennaceur}
\affiliation{Universit\'e de Lyon, F-69003 Lyon, France;
             Institut de Physique Nucl{\'e}aire de Lyon, CNRS/IN2P3,
             Universit\'e Lyon 1, F-69622 Villeurbanne Cedex,
             France}

\author{T. Duguet}
\affiliation{CEA, Irfu, SPhN, Centre de Saclay,
             F-911191 Gif-sur-Yvette,
             France}
\affiliation{National Superconducting Cyclotron Laboratory
             and Department of Physics and Astronomy,
             Michigan State University, East Lansing, MI 48824,
             USA}

\author{P.-H. Heenen}
\affiliation{PNTPM, CP229,
             Universit{\'e} Libre de Bruxelles,
             B-1050 Bruxelles,
             Belgium}

\author{T. Lesinski}
\affiliation{Universit\'e de Lyon, F-69003 Lyon, France;
             Institut de Physique Nucl{\'e}aire de Lyon, CNRS/IN2P3,
             Universit\'e Lyon 1, F-69622 Villeurbanne Cedex,
             France}
\affiliation{Department of Physics and Astronomy, 
             University of Tennessee,  
             Knoxville, TN 37996, USA}
\affiliation{Physics Division, Oak Ridge National Laboratory,
             Oak Ridge, TN 37831, USA}

\author{J. Meyer}
\affiliation{Universit\'e de Lyon, F-69003 Lyon, France;
             Institut de Physique Nucl{\'e}aire de Lyon, CNRS/IN2P3,
             Universit\'e Lyon 1, F-69622 Villeurbanne Cedex,
             France}

%
%

\begin{abstract}
We study systematically the impact of the time-even tensor terms of the
Skyrme energy density functional, i.e.\ terms bilinear in the spin-current
tensor density, on deformation properties of closed shell nuclei
corresponding to 20, 28, 40, 50, 82, and 126 neutron or proton shell
closures. We compare results obtained with three different families of
Skyrme parameterizations whose tensor terms have been adjusted on
properties of spherical nuclei: (i) $TIJ$ interactions proposed in the first
paper of this series [T.~Lesinski \emph{et al.}, Phys.\ Rev.\ C \textbf{76}, 
014312 (2007)] which were constructed through a complete readjustment of 
the rest of the
functional (ii) parameterizations whose tensor terms have been added
perturbatively to existing Skyrme interactions, with or without
readjusting the spin-orbit coupling constant. We analyse in detail the
mechanisms at play behind the impact of tensor terms on deformation
properties and how studying the latter can help screen out unrealistic
parameterizations. It is expected that findings of the present paper are
to a large extent independent of remaining deficiencies of the central
and spin-orbit interactions, and will be of great value for the
construction of future, improved energy functionals.
\end{abstract}


\date{20 September 2009}

\maketitle
%
%
\section{Introduction}
\label{sect:intro}

Our experimental knowledge of the evolution of shell structure in
atomic nuclei as a function of proton and neutron numbers has largely
increased over the last few years. The difficulty to describe these new
results has triggered the search for mechanisms that could alter nuclear
shell structure when going toward unstable nuclides and approaching nucleon
drip lines. One such mechanism that has an impact on the shell structure
of stable and unstable nuclei is provided by the tensor force
between nucleons~\cite{Ots05a}.

The tensor force is a key ingredient of all vacuum nucleon-nucleon
interactions. It is also explicitly included in the shell-model interaction
for Hamiltonians constructed from first principles. By contrast,
it was absent from methods based on the introduction of a self-consistent
mean-field~\cite{RMP}, until recent studies based on the Skyrme
or Gogny modeling of the in-medium strong
interaction~\cite{Ots06a,Bro06a,Col07a,Les07a,Zou08a,Zal08a}.

The renewed
interest in the residual tensor interaction is due to its very specific
effect on nuclear spectra.
It brings a correction to binding energies and to spin-orbit
splittings that fluctuates with the filling of shells. Its introduction
seems, therefore, necessary to improve the predictive power of mean-field-based
methods. The Skyrme and Gogny parameterizations are viewed today as nuclear
energy density functionals (EDF). Their derivation from first principles
is still lacking, although significant advances have been made
recently for the pairing part~\cite{Dug08a}. Instead, one still has to
resort to the phenomenological construction of an EDF and adjust its free
parameters, including those associated with tensor terms, to data. How to 
perform this adjustment in an optimal manner is still an unsettled question. 
The main problem is to find experimental observables which can
unambiguously be related to a mean-field result and are primarily sensitive
to a specific part of the EDF, e.g., the tensor part. Such a link between
experiment and theory is very often obscured by collective
fluctuations around the mean-field states~\cite{BBH06a,BBH08a}.

The tensor contribution to single-particle energies
depends on the filling of shells. It (nearly) vanishes in
spin-saturated nuclei, whereas it might be significant when only
one level out of two spin-orbit partners is filled.
The breaking of spherical shells in deformed nuclei leads to a strong
modification of the net spin saturation; hence, the contribution of the
tensor force to mean fields and the total energy evolves with deformation.
To the best of our knowledge, this was never studied in the published
literature. The authors of Ref.~\cite{Zal08a} considered deformation, but did
not study the impact of tensor terms on deformation properties as such.

In local energy density functionals of the Skyrme type, the tensor
force manifests itself through terms that are bilinear in the local
 pseudotensor spin-current density. In Ref.~\cite{Les07a},
referred to as Article~I in what follows, the impact of these terms on
the mean-field ground states of spherical nuclei was studied.
In the most general case, three bilinear combinations of
the pseudotensor spin-current density can be constructed from
a zero-range tensor force, one of them being also generated by
a momentum-dependent central zero-range two-body force. All of these terms
extrapolate differently when going from spherical to deformed shapes.
For the sake of simplicity, we will denote all of the three terms that are
bilinear in the spin-current tensor density as \emph{tensor terms}
throughout this article, although one of them is already present as
soon as a zero-range central velocity-dependent interaction is used 
as a starting point to derive the EDF.

A word of caution: We cannot expect to significantly improve
the agreement with all experimental data, as we learned in our study
of spherical nuclei
that most deficiencies of the single-particle spectra predicted by standard
Skyrme interactions without tensor terms persist in our fits including these
terms. Indeed, one of the major results of Article~I is that the current
form of the \emph{central} Skyrme interaction is not flexible enough to
allow for a satisfactory description of single-particle spectra.
Adding tensor terms and adjusting them tightly to very specific spectroscopic
quantifies often amplifies the deficiencies of the central and spin-orbit
parts as currently used, which seem to establish a compromise that averages
over the details of shell structure. This is consistent with the recent
study of Kortelainen \emph{et al}.~\cite{Kor08a}, who point out that it
seems impossible to satisfactorily describe single-particle levels
of doubly-magic nuclei with a standard Skyrme energy density functional
including tensor terms, even when relaxing all constraints from bulk
properties.

The aim of the present study is to investigate the generic influence of the
tensor terms on deformation properties. The questions to be addressed
here are:
\begin{itemize}
\item[(i)]
How do the tensor terms influence the topography of deformation
energy curves of even-even nuclei, given their spherical shell-structure, 
in particular regarding the number of spin-unsaturated levels at sphericity?
\item[(ii)]
How much of these changes is caused by the tensor terms
themselves, how much is caused by the rearrangement of all
other terms during the fit of the parameterizations?
\item[(iii)]
How do the three different tensor terms behave in deformed nuclei
depending on the symmetries chosen for the nuclear shapes? For
spherically symmetric systems, two of them reduce to the same functional
form, whereas the third one vanishes.
\end{itemize}
The answers to these questions will to a large extent remain independent
of remaining deficiencies of the central and spin-orbit interactions,
and will be of great value for the construction of future improved
energy functionals. We will address the question of how the surface and
surface symmetry energy coefficients change in dependence of the coupling
constants of the tensor terms, and how this influences the deformation
energy at large deformation in a future work.
%
%
\section{The Skyrme energy functional with time-reversal symmetry}
\label{sect:skyrme}

The total energy of a nucleus can be modeled by an energy density
functional~\cite{RMP,Cha97a,Cha98a,Per04a} which is the sum of five terms:
the uncorrelated kinetic energy, a Skyrme potential energy functional
that models the strong interaction in the particle-hole channel, a
pairing energy functional, a Coulomb energy functional whose
exchange term is treated using the Slater approximation and correction
terms that approximately remove the excitation energy due to spurious
motions caused by broken symmetries
\begin{equation}
\label{eq:efu:complete}
\mathcal{E}
=   \mathcal{E}_{\text{kin}}
  + \mathcal{E}_{\text{Sk}}
  + \mathcal{E}_{\text{pairing}}
  + \mathcal{E}_{\text{Coulomb}}
  + \mathcal{E}_{\text{corr}}
\, .
\end{equation}
The Skyrme energy density functional is local and can be
decomposed into isoscalar $t=0$ and isovector $t=1$ contributions
of central, spin-orbit and tensor terms
\begin{subequations}
\begin{eqnarray}
\label{eq:edf:sk}
\mathcal{E}_{\text{Sk}}
&=& \int \! d^3 r \sum_{t=0,1} \mathcal{H}^{\text{Sk}}_t  (\vec{r}) \\
&=& \int \! d^3 r \sum_{t=0,1}
  \big[   \mathcal{H}^\mathrm{c}_t  (\vec{r})
        + \mathcal{H}^\mathrm{ls}_t (\vec{r})
        + \mathcal{H}^\mathrm{t}_t  (\vec{r})
  \big]
\, .
\end{eqnarray}
\end{subequations}
The physics contained in the Skyrme functional has been discussed in
great detail in the literature~\cite{RMP,Vau72a,Bon87a,Dob00a,Per04a,Les07a}.
We will repeat here only those aspects that are relevant for the present study.
%
%
\subsection{Local densities and currents}

Each of the terms in the Skyrme energy density functional (\ref{eq:edf:sk})
can be further decomposed into one part that depends on time-even densities
only and another part that is bilinear in time-odd  densities and
currents~\cite{Per04a,Les07a}. We follow the common practice to call them 
"time-even" and "time-odd" parts of the energy density functional,
respectively, although the
energy density functional $\mathcal{E}$ itself is time-even by construction.
Throughout this article, we will assume that time-reversal symmetry is
not broken; hence, the densities and currents entering the
time-odd part of the energy density functional vanish exactly.
This allows one to represent the Skyrme part of the energy density functional
through six independent local densities:
\begin{subequations}
\begin{eqnarray}
\label{eq:locdensities:rho}
\rho_q (\vec{r})
& = & \rho_q (\vec{r},\vec{r}') \big|_{\vec{r} = \vec{r}'}
      \, ,
      \\
\label{eq:locdensities:tau}
\tau_q (\vec{r})
& = & \vnabla \cdot \vnabla' \; \rho_q (\vec{r},\vec{r}')
      \big|_{\vec{r} = \vec{r}'}
      \, ,
      \\
\label{eq:locdensities:J}
J_{q,\mu \nu}(\vec{r})
& = & - \tfrac{i}{2} (\nabla_\mu - \nabla_\mu^\prime) \;
      s_{q, \nu} (\vec{r},\vec{r}') \big|_{\vec{r} = \vec{r}'}
\end{eqnarray}
\end{subequations}
which are the scalar density $\rho_q (\vec{r})$, the scalar kinetic
density $\tau_q (\vec{r})$, and the spin-current pseudotensor density
$J_{q,\mu \nu}(\vec{r})$ for protons and neutrons $q=p$, $n$. They can be
constructed from neutron and proton density matrices expressed in the 
position basis~\cite{Dob00a,Per04a}
\begin{eqnarray}
\rho_q (\vec{r} \sigma ,\vec{r}' \sigma')
& = & \langle \hat{a}^\dagger_{r' \sigma' q} \hat{a}_{r \sigma q} \rangle
      \nn \\
& = &   \tfrac{1}{2} \, \rho_q (\vec{r},\vec{r}') \delta_{\sigma \sigma'}
      + \tfrac{1}{2} \, \vec{s}_q (\vec{r},\vec{r}')
                        \cdot \langle \sigma' | \hat{\vsigma} | \sigma \rangle
      \, ,
      \nn \\
\end{eqnarray}
where
\begin{eqnarray}
\rho_q (\vec{r}, \vec{r}')
& = & \sum_{\sigma} \rho_q (\vec{r} \sigma ,\vec{r}' \sigma) \, ,
      \nn \\
\vec{s}_q (\vec{r},\vec{r}')
& = & \sum_{\sigma \sigma'} \rho_q (\vec{r} \sigma ,\vec{r}' \sigma') \;
      \langle \sigma' | \hat{\vsigma} | \sigma \rangle
      \, .
\end{eqnarray}
Proton and neutron densities can be recoupled to isoscalar $t=0$ and
isovector $t=1$ densities, for example
$\rho_0 (\vec{r}) = \rho_n (\vec{r}) + \rho_p (\vec{r})$  and
$\rho_1 (\vec{r}) = \rho_n (\vec{r}) - \rho_p (\vec{r})$, and similarly for
$\tau_t (\vec{r})$ and $J_{t,\mu \nu} (\vec{r})$.

The cartesian spin-current pseudotensor density $J_{\mu\nu}(\vec{r})$ can be
separated into its pseudoscalar, anti-symmetric vector and
symmetric and traceless symmetric pseudotensor parts,
\begin{equation}
\label{eq:J:decomp}
J_{\mu\nu} (\vec{r})
=   \tfrac{1}{3} \delta_{\mu \nu} \, J^{(0)} (\vec{r})
  + \tfrac{1}{2} \sum_{\kappa = x}^{z} \epsilon_{\mu\nu\kappa} \,
                 J^{(1)}_{\kappa} (\vec{r})
  + J^{(2)}_{\mu\nu} (\vec{r}) \, ,
\end{equation}
where $\delta_{\mu\nu}$ is the Kronecker symbol and $\epsilon_{\mu\nu\kappa}$
the Levi-Civita tensor. The cartesian components of the pseudoscalar,
vector and traceless pseudotensor parts, expressed in terms of the cartesian
pseudotensor density, are given by
\begin{eqnarray}
\label{eq:J:recoupled}
J^{(0)}(\vec{r})
& = & \sum_{\mu = x}^{z} J_{\mu\mu}(\vec{r})\,,
       \nn \\
J^{(1)}_{\kappa} (\vec{r})
& = & \sum_{\mu,\nu = x}^{z} \epsilon_{\kappa \mu \nu} \, J_{\mu\nu}
      (\vec{r}) \,,
      \nn \\
J^{(2)}_{\mu \nu} (\vec{r})
& = &   \tfrac{1}{2} [ J_{\mu \nu}(\vec{r}) + J_{\nu \mu}(\vec{r}) ]
      - \tfrac{1}{3} \delta_{\mu\nu} \sum_{\kappa = x}^{z}
        J_{\kappa \kappa} (\vec{r})
        \,.
\end{eqnarray}
The radial component of
$\vec{J} = \sum_\kappa J^{(1)}_\kappa {\mathbf e}_\kappa$ is the only
non-zero contribution when spherical symmetry is imposed. The pseudoscalar
$J^{(0)}(\vec{r})$ term still vanishes when rotational symmetry is broken,
but parity remains conserved.
%
%
\subsection{Skyrme's tensor force}

The Skyrme energy functional representing the central, tensor, and spin-orbit
interactions can be written in different ways. The most traditional
one~\cite{Vau72a,Bon87a} is to consider the functional as generated by
a zero-range two-body effective interaction including a density-dependent
term. In his seminal articles~\cite{Sky56a,Sky58a}, Skyrme introduced two
tensor interactions that have not been considered in standard
parameterizations so far. An "even" tensor force with the coupling
constant $t_e$ mixes relative $S$ and $D$ waves, whereas an "odd" tensor
force with the coupling constant $t_o$ mixes relative $P$ and $F$ waves
\begin{eqnarray}
\label{eq:Skyrme:tensor}
v^{\text{\tensor}} (\vec{r})
& = & \tfrac{1}{2} \, t_e
    \Big\{
    \big[ 3 ( \vsigma_1 \cdot \vec{k}' )  ( \vsigma_2 \cdot \vec{k}' )
          - ( \vsigma_1 \cdot \vsigma_2 ) \vec{k}^{\prime 2}
    \big] \delta (\vec{r})
    \nn \\
& + &
     \delta (\vec{r})
      \big[ 3 ( \vsigma_1 \cdot \vec{k} ) ( \vsigma_2 \cdot \vec{k} )
            -  ( \vsigma_1 \cdot \vsigma_2)  \vec{k}^{2}
      \Big]
    \Big\}
   \nn \\
& +&  t_o
     \Big[
       3 ( \vsigma_1 \cdot \vec{k}' ) \delta (\vec{r})
          ( \vsigma_2 \cdot \vec{k} )
      - ( \vsigma_1 \cdot \vsigma_2 ) \vec{k}' \cdot
        \delta (\vec{r}) \vec{k}
     \Big] \,,
     \nn
     \\
&   &
\end{eqnarray}
where we use the shorthand notation $\vec{r} = \vec{r}_1 - \vec{r}_2$
for the relative position vector between the two particles, whereas
$\vec{k} =  - \tfrac{i}{2} ( \vnabla_1 - \vnabla_2 )$
is the operator for relative momenta acting to the right and
$\vec{k}^{\prime}$ its complex conjugate acting to the left.
The vectors formed by the Pauli spin matrices are denoted
by $\vsigma_1$ and $\vsigma_2$.

With the symmetry restrictions that we have imposed, only the
time-even part of the energy density corresponding to the tensor force
(\ref{eq:Skyrme:tensor}) is different from zero and is given by
\begin{eqnarray}
\label{eq:ef:tensor}
\mathcal{H}_t^{\text{t}}(v^{\text{\tensor}})
& = & - B^{T}_t \sum_{\mu, \nu = x}^{z} J_{t, \mu \nu} J_{t, \mu \nu}
      \nn \\
&   &
      -  B^{F}_t
         \Big[ \tfrac{1}{2} \Big( \sum_{\mu = x}^{z} J_{t,\mu \mu} \Big)2
              +\tfrac{1}{2} \sum_{\mu, \nu = x}^{z} J_{t, \mu \nu}
                                                   J_{t, \nu \mu}
         \Big]
\, .
\end{eqnarray}
The labels of the coupling constants $B^{T}_t$ and $B^{F}_t$ refer to the 
time-odd terms they multiply in the energy functional when time-reversal 
invariance is broken, ensuring Galilean invariance~\cite{Les07a}. The 
notation $\mathcal{H}_t^{\text{t}}(v^{\text{\tensor}})$ stresses that 
Eq.~(\ref{eq:ef:tensor}) provides the contribution to tensor \emph{terms}
 coming from the Skyrme zero-range tensor effective \emph{interaction} 
$v^{\text{\tensor}}$ given by Eq.~(\ref{eq:Skyrme:tensor}). When starting 
from the tensor force (\ref{eq:Skyrme:tensor}), the four
coefficients $B^{T}_t$ and $B^{F}_t$ in Eq.~(\ref{eq:ef:tensor}) are
determined by $t_e$ and $t_o$. As discussed in the next section,
the central part of Skyrme's interaction also gives rise to terms
proportional to $J_{t, \mu \nu} J_{t, \mu \nu}$ in the energy
density~\cite{Per04a,Les07a}.

The three terms present in Eq.~(\ref{eq:ef:tensor}) couple the derivatives
of the single-particle wave functions and the spin matrices in different
ways. In the first term, the derivatives contained in both $J_{t, \mu \nu}$ 
are taken along the same direction, as are the two Pauli spin matrices. The
two other terms have a structure more typical of what is expected for a
tensor interaction which couples a vector in space with a Pauli spin
matrix. In the second term  they are coupled within a given $J_{t, \mu \mu}$,
whereas in the third therm they are coupled between the $J_{t, \mu \nu}$s.
It is the simultaneous presence of these three terms that is the signature
of an actual tensor interaction.

%
%
\subsection{The Skyrme energy functional}

The complete time-even part of the Skyrme energy density functional is
obtained by combining the central, spin-orbit and tensor contributions
\begin{eqnarray}
\label{eq:EF:teven}
\mathcal{H}^{\text{Sk}}_t
& = &          C^\rho_t [\rho_0] \, \rho_t^2
             + C^{\Delta \rho}_t \rho_t \Delta \rho_t
             + C^\tau_t          \rho_t \tau_t
             + C^{\nabla \cdot J}_t \rho_t \nabla \cdot \vec{J}_t
      \nn \\
&   &        - C^{T}_t \sum_{\mu, \nu = x}^{z} J_{t, \mu \nu} J_{t, \mu \nu}
      \nn \\
&   &        - C^{F}_t \Big[ \tfrac{1}{2} \Big( \sum_{\mu = x}^{z} J_{t,\mu \mu} \Big)^2
                             + \tfrac{1}{2} \sum_{\mu, \nu = x}^{z} J_{t, \mu \nu}
                                                                    J_{t, \nu \mu}
                       \Big]
\, .
\end{eqnarray}
When the energy functional~(\ref{eq:EF:teven}) is generated from a
Skyrme interaction, the coupling constants $C_t$ are the sum of the
coupling constants $A_t$ coming from the central and spin-orbit forces
and those of the tensor force $B_t$ and are defined in the appendix~A
of Article~I.\footnote{In the expressions of $B^T_0$, $B^F_1$,
$B^{\Delta s}_1$ and $B^{\nabla s}_1$, Eqns.~(A3)-(A6) of the published
version of Article~I a global sign is missing, whereas the expressions
given in the preprint are correct.
}

One can alternatively consider $\mathcal{H}^{\text{Sk}}_t$ as
a functional of local densities in a more general sense and abandon the
link to effective interactions. The coefficients $C_t$ are then fixed
independently, except for constraints that must be imposed to fulfill
Galilean invariance, cf.~Article~I. In principle, the twelve constants
$C_t$ can furthermore depend on densities, but in all standard Skyrme
parameterizations extensively tested up to now, only $C^\rho_t$ does
depend on the isoscalar local density $\rho_0$.
%
%
\subsection{Choice of independent coupling constants in the
energy density functional}
\label{sect:choices}
The tensor terms are given in Eq.~(\ref{eq:EF:teven}) as a function
of the cartesian representation of the spin-current tensor density.
Using the pseudoscalar, vector, and pseudotensor components $J^{(0)}$,
$J^{(1)}$, and $J^{(2)}$ introduced in Eq.~(\ref{eq:J:recoupled}),
which is more appropriate when spherical symmetry is imposed,
one obtains an alternative form
\begin{eqnarray}
\label{eq:EF:tensor1}
\mathcal{H}^{\text{t}}_t
& = & - C^{T}_t \sum_{\mu, \nu = x}^{z} J_{t, \mu \nu} J_{t, \mu \nu}
      \nn \\
&   &
             - C^{F}_t \Big[ \tfrac{1}{2} \Big( \sum_{\mu = x}^{z} J_{t,\mu \mu} \Big)^2
                             + \tfrac{1}{2} \sum_{\mu, \nu = x}^{z} J_{t, \mu \nu} J_{t, \nu \mu}
                       \Big]
      \nn \\
& = &          C^{J0}_t \, \big( J_t^{(0)} \big)^2
             + C^{J1}_t \, \vec{J}_t^2
             + C^{J2}_t \sum_{\mu, \nu = x}^{z}
               J_{t, \mu \nu}^{(2)} J_{t, \mu \nu}^{(2)} \, .
      \nn \\
\end{eqnarray}
In the last line of Eq.~(\ref{eq:EF:tensor1}) we have introduced new
coupling constants for the terms bilinear in the pseudoscalar, vector and
pseudotensor parts of the spin-current pseudotensor density. Their
relation to the coupling constants defined in Eq.~(\ref{eq:EF:teven})
 is given by
\begin{subequations}
\begin{eqnarray}
C^{J0}_t
& = & - \tfrac{1}{3} C^T_t + \tfrac{2}{3} C^F_t \, ,
      \\
C^{J1}_t
& = & - \tfrac{1}{2} C^T_t + \tfrac{1}{4} C^F_t \, ,
      \\
C^{J2}_t
& = & -              C^T_t - \tfrac{1}{2} C^F_t \, .
\end{eqnarray}
\end{subequations}
In general, the tensor part of the energy density~(\ref{eq:EF:teven}) depends
on four independent parameters. This is most obvious in a cartesian
representation of the tensor functional, where these four parameters are
provided by the $C^{T}_t$ and $C^{F}_t$. In Article~I, however, we have used
two different coupling constants to characterize the tensor terms that
fulfill
\begin{subequations}
\begin{eqnarray}
C_0^{J}
& = & 2 \, C_0^{J1}
      \, ,
      \\
C_1^{J}
& = &  2 \, C_1^{J1}
      \, .
\end{eqnarray}
\end{subequations}
These two coupling constants are sufficient to describe the strength
of the isoscalar and isovector tensor terms in static\footnote{We recall
that in QRPA and other dynamical methods all components of $J_{\mu \nu}$ might
be non-zero in the response transition densities also in spherical symmetry,
as do the time-odd densities not addressed here.
}
calculations in spherical symmetry as the pseudoscalar and pseudotensor
parts of the spin-current tensor density are zero by construction.

When starting from a central and a tensor force, the ratios between the 
isospin components of the different terms will not be proportional,
i.e.\ $C_0^{J1}/C_1^{J1}$  $C_0^{J2}/C_1^{J2}$, $C_0^{T}/C_1^{T}$ and 
$C_0^{F}/C_1^{F}$ will not be equal. The
same property is lost also when one separates the tensor interaction
strength between particles of the same and different isospins.

The tensor terms of existing Skyrme parameterizations have
been adjusted on spherical nuclei, for which one has time-reversal
invariance and $\vec{J}_t^2$ is the only non-zero term in
Eq.~(\ref{eq:EF:tensor1}). Hence, the values of
$C^{J0}_t$ and $C^{J2}_t$ have not been fixed by these fits and one has
to make additional choices when going beyond sphericity.
In the present work, parity is still
conserved as a good quantum number such that
the only problem is to fix the values of the two constants $C^{J2}_t$.
The solution to this problem is not unique
and a set of reasonable choices is given by:
\begin{itemize}
\item[(i)]
One can consider that the tensor terms of the energy functional are
generated by the central and tensor parts of a Skyrme force. There is
then an univocal relation between $C^{J2}_t$ and $t_e$ and $t_o$ and
the balance between the various terms in Eq.~(\ref{eq:EF:tensor1}) is
automatically fixed. Unless otherwise noted, we will use this choice
throughout the present article for the parameter sets T$IJ$ introduced
in Article~I. This choice does not permit to set $C^{J1}_t$ and
$C^{J2}_t$ simultaneously to zero without imposing unrealistic
constraints on the central Skyrme interaction. In particular, the
parameterization T22 has been constructed in such a way that
$C^{J1}_t = 0$ and that the tensor terms vanish for spherical shapes.
The values of $C^{J2}_t$ are then non-zero and  the contribution of
the pseudotensor terms does not vanish for deformed shapes.
\item[(ii)]
Many authors set to zero the pseudotensor part of the tensor
terms~(\ref{eq:EF:tensor1}) for axially deformed Skyrme EDF calculations
\cite{Rei99a,Cha08a} although it is \emph{a priori} non-zero. They keep
only the vector part $\vec{J}$ that also appears in the spin-orbit part
of the EDF. The main motivation for this choice is that the spin-current
tensor density in cylindrical coordinates has a complicated form~\cite{Sto05a}.
This can either be viewed as an approximation or as specific choice of
the tensor terms such that $C^{J2}_t = 0$.
\item[(iii)]
Another possible choice is to set $C^F_t$ to zero. Together with suitable
choices for the time-odd part of the EDF, this allows to keep the
functional form of the standard central Skyrme EDF, but with
coupling constants of the symmetric tensor terms that are independent of 
those of a central Skyrme interaction. This choice has been made by the 
authors of Ref.~\cite{Zal08a}.
\item[(iv)]
A choice similar to the previous one is to take $C^T_t$ equal to zero,
keeping only the antisymmetric combination of the spin-current tensor density.
\item[(v)]
Finally, one can take any ratio of $C^F_t/C^T_t$ leading to a given
$C^{J1}_t$ value, which interpolates between the two previous choices.
\end{itemize}
When choosing $C^{J1}_t$ and $C^{J2}_t$ to be independent, we have
the following interrelations between coupling constants
\begin{subequations}
\begin{eqnarray}
C^{J0}_t\!\!
& = & 
        - C^{J1}_t + \tfrac{5}{6} C^{J2}_t
      \, ,
      \\
C^T_t
& = & - C^{J1}_t - \tfrac{1}{2} C^{J2}_t
       \, ,
      \\
C^F_t
& = & 2 C^{J1}_t - C^{J2}_t
\,.
\end{eqnarray}
\end{subequations}
The multitude of possible choices for the tensor pa\-ra\-me\-tri\-za\-tion
opens the risk of an inconsistent use
of the coupling constants of the tensor part of a given Skyrme parameterization.
In particular, each choice leads to very different coupling constants
in the "time-odd" part of the EDF, which can lead to significant differences.

Eventually, the bilinear part of the functional
can be recoupled into terms that contain only densities of
the same isospin on the one hand and terms that couple proton and neutron
densities on the other hand. Such a representation is often used
to characterize the interaction strength in the vector part of the
tensor terms through coupling constants $\alpha$ of the like-particle
$\vec{J}_t^2$ terms and $\beta$ of the proton-neutron $\vec{J}_t^2$ term
\begin{subequations}
\begin{eqnarray}
\label{eq:cjtot}
\alpha
& = & C_0^{J} + C_1^{J}
  =   2 ( C_0^{J1} + C_1^{J1} ) \, ,
      \\
\beta
& = & C_0^{J} - C_1^{J}
  =   2 ( C_0^{J1} - C_1^{J1} )\, .
\end{eqnarray}
\end{subequations}
The relation of $\alpha$ and $\beta$ to the coupling constants
of Skyrme's central and tensor forces can be found in Article~I.
All other coupling constants of the energy density~(\ref{eq:EF:teven})
can be recoupled in the same manner, of course.

%
%

\subsection{The single-particle Hamiltonian}

The isospin representation of the EDF is very convenient for a discussion
of its physical content. The codes that we have developed, however, use
a different representation~\cite{Bon87a} that is better suited to construct
the mean-fields with the symmetries chosen here. The central and spin-orbit
parts of the Skyrme EDF have been described in Ref.~\cite{Bon87a}. The
additional tensor terms that were not addressed are given by
\begin{eqnarray}
\label{eq:func:Skyrme:pn}
\mathcal{H}^{\text{t}}
& = &  b_{14} \sum_{\mu, \nu = x}^{z} J_{0, \mu \nu} J_{0, \mu \nu}
       \nn \\
&   & 
             + b_{16} \Big[  \Big( \sum_{\mu = x}^{z} J_{0, \mu \mu} \Big)^2
                           + \sum_{\mu, \nu = x}^{z} J_{0, \mu \nu} J_{0, \nu \mu}
                      \Big]
       \nn \\
&   &  + \sum_{q=n,p} \Big\{
               b_{15} \sum_{\mu, \nu = x}^{z} J_{q,\mu \nu} J_{q,\mu \nu}
       \nn \\
&   & 
             + b_{17} \Big[ \Big(  \sum_{\mu = x}^{z} J_{q,\mu \mu} \Big)^2
                           + \sum_{\mu, \nu = x}^{z} J_{q, \mu \nu} J_{q, \nu \mu}
                       \Big)
             \Big\}
\, .
\end{eqnarray}
The coupling constants of~(\ref{eq:func:Skyrme:pn}) are related to those
of~(\ref{eq:EF:tensor1}) through
\begin{eqnarray}
b_{14}
& = & - C^T_0 + C^T_1
  =     C^{J1}_0 - C^{J1}_1 + \tfrac{1}{2} C^{J2}_0 - \tfrac{1}{2} C^{J2}_1
      \nn \\
b_{15}
& = & - 2 C^T_1
  =     2 C^{J1}_1 + C^{J2}_1
      \nn \\
b_{16}
& = & - \tfrac{1}{2} \, C^F_0
      + \tfrac{1}{2} \, C^F_1
  =   - C^{J1}_0
      + C^{J1}_1
      + \tfrac{1}{2} \, C^{J2}_0
      - \tfrac{1}{2} \, C^{J2}_1
      \nn \\
b_{17}
& = & - C^F_1
  =   - 2 C^{J1}_1 + C^{J2}_1
\, .
\end{eqnarray}
The mean-field equations for protons and neutrons, obtained by
functional derivative techniques~\cite{RMP,Per04a} from the energy
functional~(\ref{eq:func:Skyrme:pn}), read
\begin{equation}
\label{eq:mf:eq}
\hat{h}_q (\vec{r})\, \psi_i(\vec{r})
= \epsilon_i \, \psi_i (\vec{r})
\, ,
\end{equation}
with the one-body Hamiltonian corresponding to the energy
functional~(\ref{eq:EF:teven}) given by\footnote{For the standard
Skyrme functional~(\ref{eq:EF:tensor1}) with non-density-dependent
coupling constants of the spin-orbit and tensor terms,
the second line can be simplified into
$\displaystyle -i \sum_{\mu\nu}W_{q,\mu\nu}(\vec{r})\nabla_\mu \hat\sigma_\nu$
for the symmetries chosen here.
}
\begin{eqnarray}
\label{eq:sphamil}
\hat{h}_q (\vec{r})
& = &   U_q (\vec{r})
      - \vnabla \cdot B_q (\vec{r}) \vnabla
      \nn \\
&   &
      - \tfrac{i}{2}
        \sum_{\mu, \nu = x}^{z}
        \big[ W_{q, \mu \nu} (\vec{r}) \, \nabla_\mu
              + \nabla_\mu \, W_{q, \mu \nu} (\vec{r})
        \big] \, \hat{\sigma}_\nu
\, ,
\end{eqnarray}
where the $\hat{\sigma}_\nu$ denote the Pauli matrices.
The expressions for the single-particle potential $U(\vec{r})$ and the inverse
effective mass $B(\vec{r})$ are the same as those given in Ref.~\cite{Bon87a}.
When the energy functional depends on the vector part of the spin-current
tensor only, the second line of Eq.~(\ref{eq:sphamil}) boils down to
$-i \vec{W}_q (\vec{r}) \cdot \vnabla \times \vsigma$ with
$\vec{W}_q (\vec{r}) = \sum_{\mu \nu \kappa} \epsilon_{\mu \nu \kappa}
W_{q,\mu \nu} (\vec{r}) \, \vec{e}_\kappa$, where $\vec{e}_\kappa$ is the
unit vector in $\kappa$ direction. With the full spin-current tensor, one
has to consider

\begin{eqnarray}
\label{eq:Wmunu}
\lefteqn{
W_{q,\mu \nu} (\vec{r})
} \nn \\
& = & \frac{\delta E}{\delta J_{q,\mu \nu} (\vec{r})}
      \nn \\
& = & -  b_{9} \sum_{\kappa=x}^{z} \epsilon_{ \kappa \mu \nu}
        \big(  \nabla_\kappa \rho
             + \nabla_\kappa \rho_q
        \big)
      \nn \\
&   & + 2 \, b_{14} \, J_{\mu \nu}
      + 2 \, b_{15} \, J_{q,\mu \nu}
      + 2 \, b_{16} \, J_{\nu \mu}
      + 2 \, b_{17} \, J_{q,\nu \mu}
      \nn \\
&   &
      + 2 \, b_{16} \, \Big[ \sum_{\kappa=x}^{z} J_{\kappa \kappa} \Big]
        \delta_{\mu \nu}
      + 2 \, b_{17} \, \Big[ \sum_{\kappa=x}^{z} J_{q, \kappa \kappa} \Big]
        \delta_{\mu \nu}
\end{eqnarray}
instead. The terms in the first line of Eq.~(\ref{eq:Wmunu}) originate from
the spin-orbit part of the functional, the other two lines from the tensor
part.  The terms in the last line of Eq.~(\ref{eq:Wmunu}) might be nonzero
only when parity is broken.

We recall that for
constrained calculations, as discussed below, the constraints do not
contribute to the observable total energy, which is still obtained from
$\mathcal{E}$, Eq.~(\ref{eq:efu:complete}), whereas the eigenvalues
$\epsilon_i$ of the mean-field Hamiltonian used to construct the Nilsson
diagrams contain a contribution from the constraint~\cite{aberg90}.

%
%
\section{Parameterizations}
\label{sect:param}

For an overview of earlier choices made for the coupling constants of the
tensor terms in standard parameterizations of the Skyrme energy functional,
we refer to Article~I. We will limit ourselves here to recent parameterizations
that explore the impact of tensor terms on single-particle spectra.

%
%
\subsection{T$IJ$ parameterizations of Lesinski \emph{et al}}

\begin{figure}[t!]
\centerline{\includegraphics{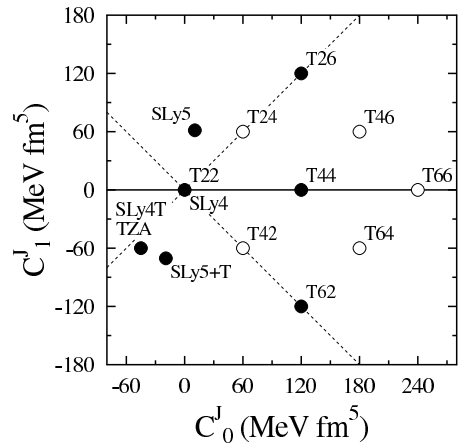}}
\caption{
\label{fig:coupling}
Coupling constants $C^J_0$ and $C^J_1$ of the tensor terms for the
parameterizations discussed in the paper.
}
\end{figure}

The main aim of the present article is to test the deformation
properties of magic and semi-magic nuclei obtained with the
parameterizations T$IJ$ introduced in Article~I. The fit of these
parameterizations is based on the same protocol as the one of the
SLy$x$ parameterizations~\cite{Cha97a,Cha98a},
with a few minor changes explained in Article~I. We have found
in Article~I that to add a tensor term to a standard Skyrme EDF
does not globally correct its deficiencies for the prediction of
masses, radii, or single-particle properties of semi-magic nuclei.
In fact, very different values of the $C^J_t$ constants are required
for each of these observables to be accurately reproduced, and even these
values should vary in different mass regions. Instead of trying to construct
a single "best" EDF in Article~I, we have studied  the impact of the
tensor terms on different observables by constructing a set of 36
parameterizations, each corresponding to given values of  $C^J_0$ and $C^J_1$,
and all other Skyrme parameters being determined by the same fitting procedure.
In this way, a wide range of the effective coupling constants $C^J_0$
and $C^J_1$ is systematically covered.

In the present study, we limit ourselves to a small subset of these
36 parameterizations, i.e.\  T22, T26, T44, and T62 in most cases.
Figure~\ref{fig:coupling} shows their location in the parameter
space of $C^J_0$ and $C^J_1$. The parameterization T22 has by construction
vanishing tensor terms at sphericity. It should have properties close to
those of SLy4 in which tensor terms have been neglected.
The parameterizations T26, T44 and T62 share the same value of the
isoscalar tensor coupling constant $C^J_0 = 120$ MeV fm$^5$,
and differ by the isovector one $C^J_1$, which takes
the values 120, 0 and $-120$~MeV~fm$^5$ respectively.

Parameterizations having the same proton-neutron coupling constant $\beta$
are aligned along the first diagonal in Fig.~\ref{fig:coupling},
those with the same value of $\alpha$ are aligned along the
anti-diagonal. For parameterizations T$IJ$, the coefficient of the
proton-neutron tensor term increases with the first index $I$ for the fixed
like-particle tensor term, whereas that of the like-particle tensor term
increases with the second index $J$ for fixed proton-neutron tensor
coupling. Let us recall that the integers $I$ and $J$
are related to the constants $\alpha$ and $\beta$ by
\begin{subequations}
\begin{eqnarray}
\alpha 
& = & 60(J-2) \; \text{MeV} \; \text{fm}^5 , \\
\beta 
& = & 60(I-2) \; \text{MeV} \; \text{fm}^5 \, .
\end{eqnarray}
\end{subequations}

\begin{table*}[ht]
\caption{
\label{tab:interaction}
Skyrme parameterizations discussed in this work. Procedure:
Variational (V) corresponds to Skyrme parameterizations where, for given tensor
coefficients, all other Skyrme parameters are fitted following the procedure
of Ref.~\cite{Cha98a}, slightly modified in Article~I for the T$IJ$ 
interactions, while perturbative (P) labels parameterizations for which 
tensor terms are added without refit. Type: the parameterization is 
treated as an interaction (I) or a functional (F). Central: the contribution 
to the tensor terms coming from the central part of the interaction is 
included or not. Tensor: a tensor interaction is included or not
}
\begin{tabular}{lccccc@{\hspace{4mm}}p{8.5cm}}
\hline\noalign{\smallskip}
           & procedure &  type  &  central & tensor & Ref. & remarks  \\
\noalign{\smallskip}\hline\noalign{\smallskip}
SLy4         &   V &  I &  N   & N & \cite{Cha98a} & The tensor contributions from the central part of the interaction are neglected.\\
SLy5         &   V &  I &  Y   & N & \cite{Cha98a}&  \\
T$IJ$        &   V &  I &  Y   & Y & \cite{Les07a} & Isovector tensor coefficients equal to zero if $I=J$ \\
T22          &   V &  I &  Y   & Y & \cite{Les07a} &
Central and tensor contributions to the tensor terms such that they cancel each other at sphericity (close to SLy4)\\
\noalign{\smallskip}\hline\noalign{\smallskip}
SLy4T         &   P &  F &  -   & - &\cite{Zal08a}  &\\
SLy4T$_{\text{min}}$   &  V  &  F &  -   & - &\cite{Zal08a}  & Refit of SLy4T on masses keeping tensor and spin-orbit coefficients fixed\\
SLy5+T        &   P &  I &  Y   & Y & \cite{Col07a} &\\
\noalign{\smallskip}\hline\noalign{\smallskip}
SLy4T$_{\text{self}}$ & V &  I &   Y  & Y & this work &  Refit of SLy4T with the same protocol as T$IJ$ keeping the same spin-orbit and tensor coefficients for spherical shapes as in Ref.~\cite{Zal08a} \\
TZA           &   V &  I &   Y  & Y & this work  & Refit of SLy4T with the same protocol as T$IJ$ for the tensor coefficients used in Ref.~\cite{Zal08a}\\
\noalign{\smallskip}\hline
\end{tabular}
\end{table*}

%
%
\subsection{The parameterization of Col{\`o} \emph{et al}.}

The Skyrme parameterization SLy5 introduced in Ref.~\cite{Cha98a} is one
of the two SLy$x$ functionals which include the tensor terms generated
from the central part of the Skyrme force.
Col{\`o} \emph{et al.}~\cite{Col07a} have added a tensor force
to it, keeping all the other coupling constants of the parameterization
at their original values. We will call this interaction SLy5+T
("SLy5 plus tensor") in what follows. The
parameters of the tensor force were adjusted in spherical symmetry to
single-particle energies along the chains of $N=82$ isotones and $Z=50$
isotopes. The empirical values for these energies were obtained as separation
energies of the last particle in states of an odd-$A$ nucleus, assumed to
be dominated by one single-particle configuration. They were compared to the
eigenvalues of the one-body Hamiltonian, Eq.~(\ref{eq:sphamil}), in the
neighboring even-even nucleus.
The resulting coupling constants in MeV fm$^5$ are $C^J_0 = -19.333$ and
$C^J_1 = -70.466$, or, equivalently, $\alpha = -89.8$ and $\beta = 51.9$.
This parameterization has been used in studies of spherical shell structure
in Ref.~\cite{Zou08a} and of the Gamov-Teller strength
distribution in \nuc{90}{Zr} and \nuc{208}{Pb} through RPA calculations
in Ref.~\cite{Bai09a,Bai09b}. As can be seen in Fig.~\ref{fig:coupling},
SLy5+T explores a different region of the parameter space than the T$IJ$
parameterizations; $\alpha$ being negative and its modulus larger than $\beta$.

%
%
\subsection{The parameterization of Zalewski \emph{et al}.}

In Ref.~\cite{Zal08a}, Zalewski \etal\ did refit the spin-orbit and
tensor coupling constants of some standard Skyrme interactions.
We will consider here two of their fits that are based on
the SLy4 functional~\cite{Cha98a}.
In a first step, Zalewski \etal\ readjusted $C^{\nabla J}_t$ and $C^{J}_t$
to carefully selected spin-orbit splittings in \nuc{40}{Ca}, \nuc{48}{Ca}
and \nuc{56}{Ni}, keeping all other coupling constants of the energy
functional at their original values.
The single-particle energies of a spherical mean-field calculation of
doubly-magic nuclei have been identified with the separation energy of a
nucleon with the same quantum numbers to or from the odd neighboring nuclei.
They were compared to experimental separation energies corrected through
a macroscopic model taking into account the influence
of the coupling of the single-particle state to collective vibrations
of the surface. The  values of the parameters
resulting from this procedure are $C^J_0 = -45$, $C^J_1 = -60$,
$C^{\nabla J}_0 = -60$ and $C^{\nabla J}_1 = -20$ (all in MeV fm$^5$)
and define an EDF called SLy4T. The strength of the standard zero-range
spin-orbit force of SLy4T is equal to $W_0 = 80$ MeV fm$^5$ and is much
lower than in the original SLy4 parameterization, for which
$W_0 = 123$~MeV fm$^5$.

Such a modification of  $C^J_t$ and $C^{\nabla J}_t$ from their original values
without changing the other parameters of the EDF degrades prohibitively
the masses calculated with SLy4T with respect to those obtained with SLy4.
For this reason, Zalewski \etal\ refitted all parameters of SLy4T except
$C^J_t$ and $C^{\nabla J}_t$ in a second step to restore a reasonable
description of bulk properties, leading to the parameterization
SLy4T$_{\text{min}}$.

The parameterizations SLy4T and SLy4T$_{\text{min}}$ are explicitly
constructed as energy density functionals  without making reference
to any underlying central, spin-orbit, or tensor force.
In particular, the authors chose
to set the two coupling constants $C^F_t$ of the asymmetric
cartesian tensor term to zero and to vary only
$C^T_t$. This automatically fixes the value of $C^{J2}_t$ to
be equal to two times that of $C^{J1}_t$ for $t=0$ and $t=1$.
Although the coupling constants $C^{\nabla J}_t$ of the spin-orbit term
were readjusted, the ratio between the isoscalar and isovector coupling
constants $C^{\nabla J}_0 / C^{\nabla J}_1$ was kept at the value of the
original fit.

To analyze the consequences of the choices made in the fitting strategy
of Zalewski \etal\ and those of Article~I, we performed two
additional fits using the same protocol as in Ref.~\cite{Les07a}, but
exploring a different region around
($-45$, $-60$)~MeV\,fm$^5$ corresponding to SLy4T in the
$C^J_0$, $C^J_1$ plane of  Fig.~\ref{fig:coupling}. For the first one,
called SLy4T$_{\text{self}}$ hereafter,
we fixed  $C^J_t$ and $C^{\nabla J}_t$ at their SLy4T values, but readjusted
all the other constants of the functional to obtain the "best" EDF
corresponding to our protocol. This parameterization differs from
SLy4T$_{\text{min}}$ by the fit protocol, and by our choice
to keep the interrelations between the coupling constants of the
tensor terms as obtained from a two-body central and tensor forces.
To study also the impact of the readjustment of the spin-orbit interaction,
we constructed a second parameterization, called TZA hereafter,
where we additionally vary $W_0$, resulting to a value
$W_0 = 111.934$~MeV\,fm$^5$. This parameterization is thus fitted exactly as
the T$IJ$ ones, except that it is outside of the rectangular parameter
space for $C^{J}_0$ and $C^{J}_1$ considered in Article~I.
The coupling constants for SLy4T$_{\text{self}}$ and TZA can be found in
the \emph{Physical Review} archive \cite{EPAPS}. The properties of all the interactions
that we have used are summarized in Table~\ref{tab:interaction}.

%
%
\section{Results}
\label{sect:results}

\subsection{Technical Details}

The wave functions  are constructed with the code EV8~\cite{Bon85a,Bon05a}
which has been modified to include the tensor terms in the
energy density functional and the single-particle Hamiltonian.
The energies have been recalculated after convergence with a code that
uses a more accurate algorithm for the derivatives.

Pairing correlations are treated with the
Lipkin-Nogami (LN) method to avoid the breakdown of BCS pairing and
the resulting discontinuities in the deformation energy curves.
We use an effective density-dependent zero-range pairing interaction
with two soft cutoffs at 5 MeV above and below the Fermi energy as
described in Ref.~\cite{Rig99a}. For consistency with our recent
calculations~\cite{Ben03a,BBDH04a}, we chose a strength of
$-1000$ MeV fm$^{-3}$ for light and medium-heavy nuclei and
$-1250$ MeV fm$^{-3}$ for \nuc{186}{Pb} and \nuc{208}{Pb}.

%
%
\subsection{General comments}

The magic numbers close to stability can be divided into
two categories: up to 20, they correspond to a spin-saturated
closure of major oscillator shells, whereas above 20, they are created by
the spin-orbit interaction which pushes down the level with largest $j$-value
into the gap between the oscillator shells for 28, or even into the oscillator
shell below the gap for 50, 82 and 126. One usually labels a spherical
nucleus spin saturated when all pairs of spin-orbit partners are either
occupied or empty. In realistic mean-field calculations this will not
result in an exact cancellation of the spin-current tensor density, as it
should be the case for an exact spin saturation. This has two origins.
First, the radial wave functions of the spin-orbit partners are not identical,
and, second, pairing correlations will smear out the distribution
of occupation numbers. The effect of these two factors will be discussed
in a forthcoming publication~\cite{Ben09y}.
Tensor terms in the time-even part of the
energy functional fluctuate with the spin-current tensor density
$J_{\mu \nu}$, which is small in  spin-saturated systems, and large
whenever only the lower level of a pair of spin-orbit partners is filled,
while its partner level remains empty. For an illustration we refer to
Article~I.

Deformation breaks this simple picture. As soon as it sets in, the spin
saturation or non-saturation disappears and the energy due to tensor
terms varies in a way related to the sign of the coupling constants.
Close to sphericity, this can be determined by looking at the $C^J_t$
coefficients.
For $N=Z$ spin saturated nuclei and parameterizations with $C^J_0$ positive,
the contribution from the tensor interaction is zero at sphericity and
becomes repulsive as soon as deformation sets in. For $N=Z$ spin-unsaturated
nuclei, the tensor contribution will be largest at sphericity and
decrease with deformation.

%
%
\subsection{\nuc{56}{Ni}}

\begin{figure}[htbp]
\centerline{\includegraphics{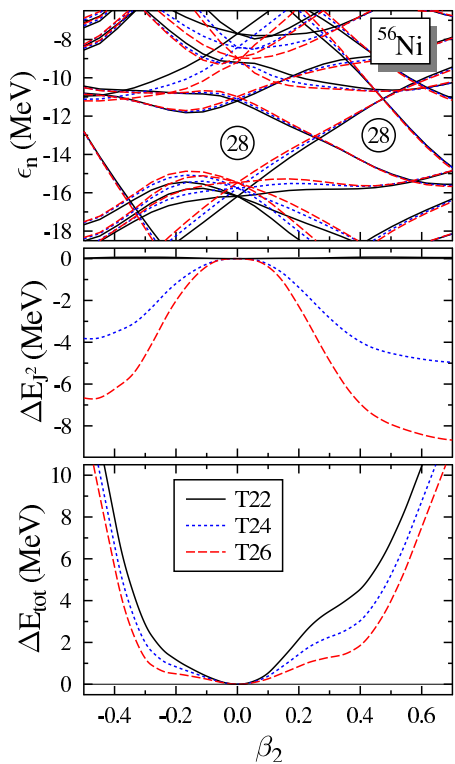}}
\caption{
\label{fig:ni56:trends}
(Color online)
Nilsson diagram of the neutrons (top), change of the total contribution
from the tensor terms to the total energy relative to the values at
the spherical shape (middle) and deformation energy relative to the
spherical shape (bottom) for \nuc{56}{Ni} obtained with the
parameterizations T22, T24 and T26. The energy scale
is the same for the two lower panels.
}
\end{figure}

Let us start our study by looking in detail into \nuc{56}{Ni},
the lightest doubly-magic nucleus with major proton and neutron
shell closures due to the spin-orbit interaction. The values and
systematics of static~\cite{Ber09aE} and transition moments~\cite{Orc08aE}
of low-lying states around \nuc{56}{Ni} suggest that it is not as good an
inert magic core as other doubly-magic nuclei. This feature is also observed
in shell-model calculations~\cite{Ots98a} and is at the origin of substantial
corrections found between "empirical" and "bare" single-particle
energies in Ref.~\cite{Tra96aE}. Several well-deformed rotational
bands coexisting with the spherical shell-model-type states have been
observed, one of them down to a $2^+$ level at 5.351 MeV~\cite{Ru99a}.

%
%
\subsubsection{Key quantities}

Figure~\ref{fig:ni56:trends} provides three key quantities for the analysis of
the deformation properties of \nuc{56}{Ni}: the neutron Nilsson diagram,
the contribution of tensor terms to the deformation energy  and the total
energy, all as a function of axial quadrupole deformation. Three
parameterizations have been used, T22, T24 and T26, which differ in the
strength of the tensor terms. The proton Nilsson
diagram is very similar to the one for neutrons, except for
an overall shift due to the Coulomb interaction.
The dependence of these quantities on the axial quadrupole deformation is shown
as a function of the dimensionless deformation $\beta_2$ of the mass density
distribution defined as
\begin{equation}
\label{eq:beta2}
\beta_2
= \sqrt{\frac{5}{16\pi}} \, \frac{4\pi}{3 R^2 A} \,
  \langle 2 z^2 - y^2 - x^2 \rangle
\, ,
\end{equation}
where $R = 1.2 \, A^{1/3}$ fm.

The presence of tensor terms in the energy
functional has an obvious impact on the single-particle levels. An increase
of the tensor interaction results in a reduction of the spin-orbit
fields and in a smaller spherical gap at $N=28$. The net result is
a sizable decrease of the splitting of the $1f$ levels from T22 to T62.
At the same time, tensor terms also modify the slope of the Nilsson levels
at small deformations, whereas, at large deformation, the levels predicted
by the three parameterizations nearly lie on top of each other.

At sphericity, the tensor contribution for T22 is zero by construction.
In practice, one can see that the tensor energy remains close to zero
for all deformations. As soon as the nucleus is deformed, the decrease
of spin non-saturation
strongly affects the tensor terms. Parameterizations
like T24 and T26 have a positive like-particle coupling constant
$\alpha$ and give repulsive tensor energies at sphericity. This repulsion
is decreased by deformation, which reduces the tensor terms by several MeV.
For T26, the total energy curve
obtained as a function of deformation is softer than without the tensor
interaction. In particular, the shoulder at prolate deformations becomes
lower in energy and more pronounced. This structure is associated with the
rotational band observed down to spin $2^+$~\cite{Ru99a}. The gain in
total deformation energy, however, is much smaller than the gain in
deformation energy from the tensor terms.

In the following subsections, we will analyze the origin of these
differences and their dependence on the fit strategy.

%
\subsubsection{Contributions to the total energy}

Let us first recall that the interactions constructed in Article~I differ
not only by their choice of the strengths of the tensor interaction, but
that all
the other terms of the energy functional are also different because each
interaction is refitted on the same set of data. It is, therefore, interesting
to examine how the different terms of the functional vary from one set to
another and how the changes induced by the tensor interaction are, in fact,
largely attenuated by a readjustment of the entire functional.

\begin{figure}[t!]
\centerline{\includegraphics{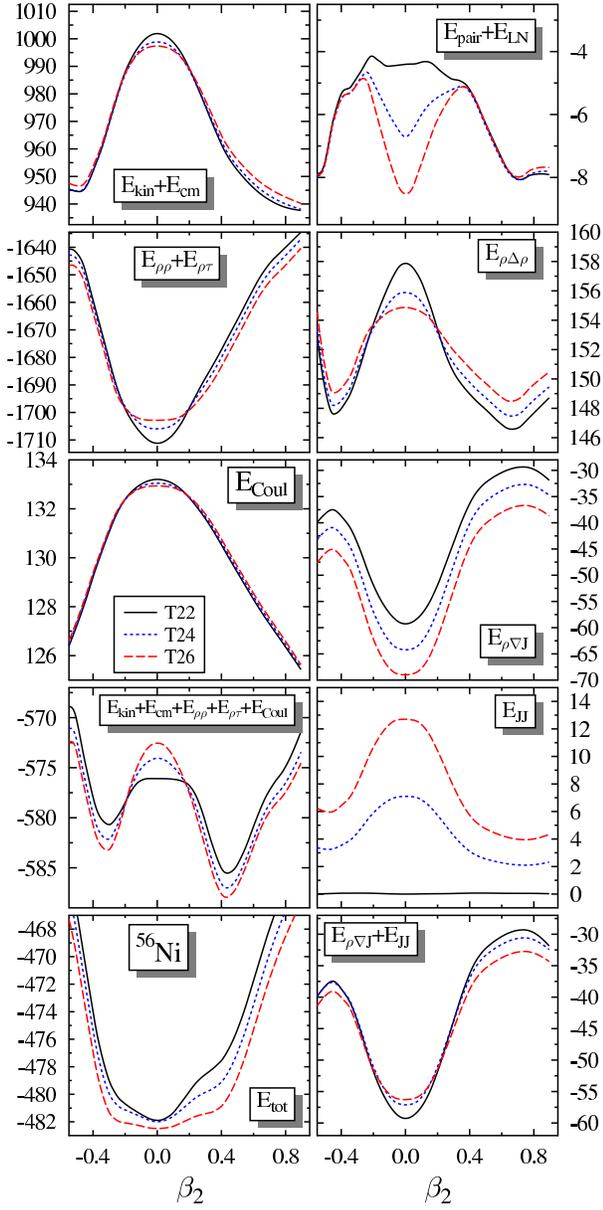}}
\caption{
\label{fig:ni56:total:decomposition}
(Color online)
Decomposition of the total energy of \nuc{56}{Ni} obtained
with the parameterizations T22, T24 and T26 into the various
contributions to the EDF, Eqns.~(\ref{eq:efu:complete})
and~(\ref{eq:EF:teven}), as a function of the quadrupole
deformation $\beta_2$ (see text).
}
\end{figure}

\begin{figure}[t!]
\centerline{\includegraphics{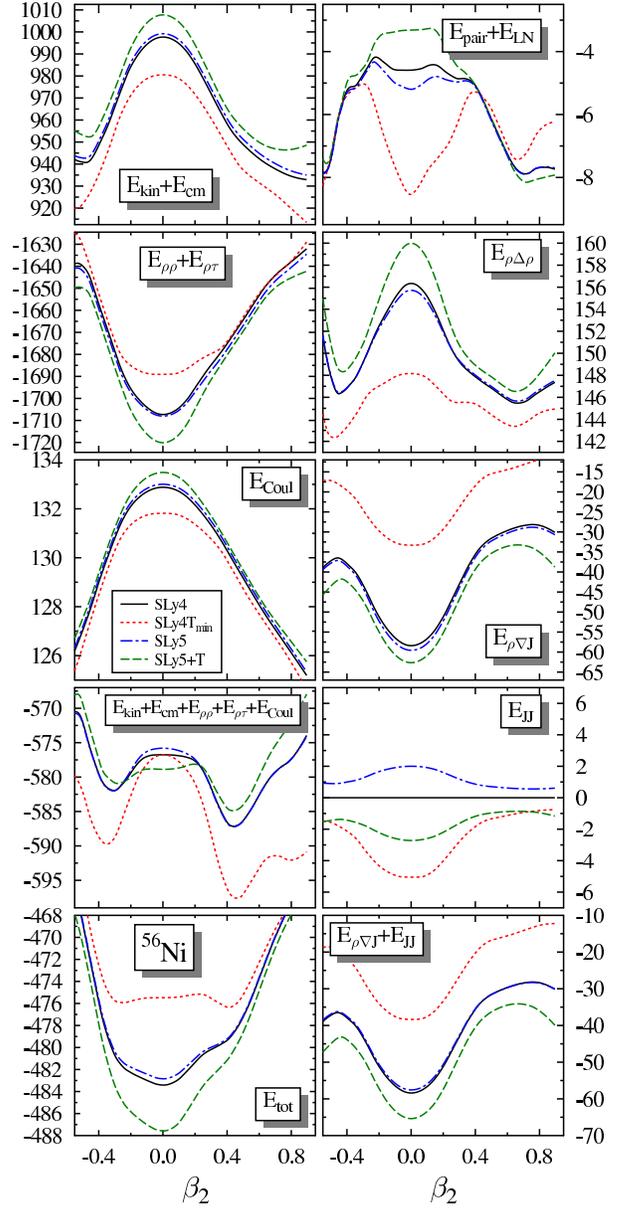}}
\caption{
\label{fig:ni56:total:decomposition2}
(Color online)
Same as Fig.~\ref{fig:ni56:total:decomposition}, but for
the parameterizations SLy4, SLy4T$_{\text{min}}$, SLy5, and SLy5+T.
}
\end{figure}

Figure~\ref{fig:ni56:total:decomposition} presents the decomposition of the
total binding energy into the contributions from the various terms in the
energy functional, Eq.~(\ref{eq:efu:complete}), for parameterizations
T22, T24 and T26. Those panels showing contributions to
the Skyrme energy functional are labeled by their content in densities,
Eq.~(\ref{eq:EF:teven}), whereas the other panels provide the kinetic 
energy (plus the one-body center-of-mass correction), the Coulomb energy 
and the pairing energy (including the Lipkin-Nogami correction).

Unlike in Fig.~\ref{fig:ni56:trends},
Fig.~\ref{fig:ni56:total:decomposition} shows here the absolute
values for the total binding energy and the tensor contributions.
The binding energy of \nuc{56}{Ni}  is included in the data
the T$IJ$ parameterizations are adjusted to and, indeed, the total
energy (lower left panel) differs by only a few 100 keV at sphericity.
Also, as already pointed out, the deformation energy curves obtained
with the three parameterizations differ by less than 2 MeV. These
similarities
result from a complicated compensation between the various components
of the energy. All of them, with the exception of the Coulomb energy,
differ on a much larger scale, both in absolute values and deformation
dependence.

It can be seen also that each term of the EDF has a very different dependence
on deformation and that the total deformation energy also always results from
subtle compensations. Both the kinetic energy and the part of the Skyrme
functional that contributes to $E/A$ in infinite homogeneous nuclear matter,
$C^\rho[\rho] \rho^2 + C^\tau \rho \tau$, vary by about 70 MeV as a function
of deformation. The part of the EDF that does not depend on gradient terms
is obtained by summing these two terms and the Coulomb energy, and is
represented
in the panel in the fourth row on the left. It varies with deformation by
about 15~MeV. For all parameterizations, the latter curves exhibit pronounced
prolate and oblate minima. The gradient term $C^{\Delta \rho} \rho \Delta \rho$
even amplifies the preference for deformed minima. The combined spin-orbit
and tensor terms, shown individually and summed up in the three lower right
panels, are the necessary ingredient to obtain a spherical ground
state in \nuc{56}{Ni}. This underlines the fact that the spin-orbit and tensor
terms are not only important for
single-particle spectra, but also might play a crucial role
for the total binding energy, in particular for its deformation dependence.
Interestingly, for this nucleus and the parameterizations shown, the
contributions of the spin-orbit and tensor terms are of opposite sign
and sum up such that their sum is much less dependent on the parameterization
than the individual terms.

Figure~\ref{fig:ni56:total:decomposition2} providess the same decomposition
for various variants of SLy4 adjusted with different strategies,
i.e.\ SLy4, SLy5, SLy4T$_{\text{min}}$ and SLy5+T.

The results obtained with the two functionals SLy4 and SLy5 adjusted in
Ref.~\cite{Cha98a} are quite close for all terms, except of course
for the tensor contribution, excluded in the case of SLy4 and restricted
to the contribution from the central interaction for SLy5. All components
of the energy differ slightly since the coupling constants are completely
refitted in both cases. The total energies obtained with these two
parameterizations, however, are quite close.

In particular, although SLy4 and SLy5 correspond to slightly different
coupling constants, they lead to results which differ on a much smaller
scale than SLy5 and SLy5+T, which differ only by the tensor terms. This
underlines the role of self-consistency: the difference in the tensor
terms is responsible for changes in the single-particle properties,
which ultimately induce changes in each individual contribution to the
energy functional.

The similarity between the curves obtained with SLy4 and T22
(Fig.~\ref{fig:ni56:total:decomposition} and
Fig.~\ref{fig:ni56:total:decomposition2}), shows that the slight differences
between the interactions have no significant effect.

\begin{figure}[t!]
\centerline{\includegraphics{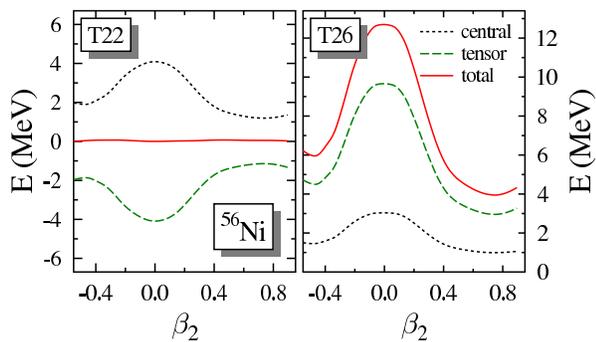}}
\caption{
\label{fig:ni56:tensor:decomposition}
(Color online)
Decomposition of the isoscalar tensor energy obtained
in \nuc{56}{Ni} for the parameterizations T22 and T26
into the
contributions from the central and tensor parts of the forces.
}
\end{figure}

%
%
\subsubsection{Decomposition of the tensor terms}

The energy contribution from the tensor terms can be decomposed in several
ways. We first compare the contributions from the central and tensor parts
of the parameterizations T22 and T26 in
Fig.~\ref{fig:ni56:tensor:decomposition}. As explained in
Sect.~\ref{sect:choices}, such a decomposition has a meaning only when
assuming an underlying force, but not for genuine functionals. The central
contribution  is very similar for both parameterizations (and all others
from the T$IJ$ family),
which is a consequence of its correlation with effective masses and surface
tensions through $t_1$ and $t_2$ terms of the two-body Skyrme force,
see the discussion of Fig.~3 in Article~I. The contribution from central
and tensor parts cancel nearly exactly for T22 for all deformations. As
exemplified by T26, the contributions from the central and tensor forces
to the tensor terms have the same sign for \nuc{56}{Ni} for all other T$IJ$
parameterizations
that have zero or positive values for $C^J_0$ and $C^J_1$.

\begin{figure}[t!]
\centerline{\includegraphics{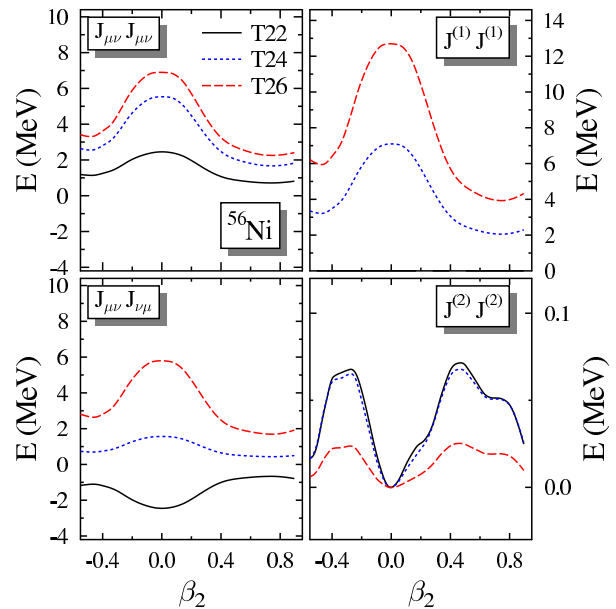}}
\caption{
\label{fig:ni56:tensor:decomposition2}
(Color online)
Decomposition of the isoscalar tensor energy obtained in \nuc{56}{Ni}
for the parameterizations T22, T24 and T26 into the contributions from
symmetric and asymmetric terms in the cartesian representation (left),
or the contributions from vector and pseudotensor contributions (right).
The same energy scale is used for all panels except the one for the pseudotensor contribution
$J^{(2)}J^{(2)}$.
}
\end{figure}

\begin{table*}[ht!]
\caption{
\label{tab:ni56}
Eigenvalues corresponding to the neutron $pf$ and the $g_{9/2^+}$ orbitals
obtained for the single-particle Hamiltonian at
spherical shape in \nuc{56}{Ni} (see text).
}
\begin{tabular}{lccccccccccc}
\hline\noalign{\smallskip}
parameterization &
$\epsilon_{f_{7/2}}$ &
$\epsilon_{f_{5/2}}$ &
$\Delta \epsilon_f$  &
$\epsilon^{\text{cent}}_f$ &
$\epsilon_{p_{3/2}}$ &
$\epsilon_{p_{1/2}}$ &
$\Delta \epsilon_p$ &
$\epsilon^{\text{cent}}_p$ &
$\Delta \epsilon_f/\Delta \epsilon_p$ &
$\epsilon^{\text{cent}}_f - \epsilon^{\text{cent}}_p$ &
$\epsilon_{g_{9/2}}$ \\
\noalign{\smallskip}\hline\noalign{\smallskip}
experiment  ($\pm S_{1n}$) & -16.64 & -9.48 & 1.02 & -13.57 & -10.25 & -9.14 & 0.37 &  -9.88 & 2.75 & 3.69 &  --   \\
empirical \cite{Tra96aE}   & -16.93 & -9.53 & 1.06 & -13.76 & -10.36 & -8.48 & 0.63 &  -9.73 & 1.68 & 4.03 &  --   \\
\noalign{\smallskip}\hline\noalign{\smallskip}
T22             &   -16.18  & -7.70 & 1.21 & -12.54 &   -11.21 & -9.19 & 0.67 & -10.54  &    1.80 &  2.00 &    -4.26 \\
T26             &   -15.47  & -8.97 & 0.93 & -12.68 &   -10.98 & -9.19 & 0.60 & -10.38  &    1.56 &  2.30 &    -3.56 \\
T44             &   -15.56  & -8.69 & 0.98 & -12.61 &   -11.09 & -9.20 & 0.63 & -10.46  &    1.56 &  2.15 &    -3.75 \\
T62             &   -15.61  & -8.52 & 1.01 & -12.57 &   -11.22 & -9.25 & 0.66 & -10.56  &    1.54 &  2.00 &    -3.89 \\
\noalign{\smallskip}\hline\noalign{\smallskip}
SLy5            &   -16.01  & -8.03 & 1.14 & -12.59 &   -11.11 & -9.17 & 0.65 & -10.47  &    1.76 &  2.12 &    -4.05 \\
SLy5+T          &   -16.66  & -7.09 & 1.37 & -12.56 &   -11.15 & -9.01 & 0.72 & -10.44  &    1.91 &  2.12 &    -4.67 \\
\noalign{\smallskip}\hline\noalign{\smallskip}
SLy4                       & -16.17 & -7.80 & 1.20 & -12.58 & -11.13 & -9.14 & 0.66 & -10.47 & 1.81 & 2.11 & -4.20 \\
SLy4T                      & -15.49 & -8.71 & 0.97 & -12.58 & -11.20 & -9.47 & 0.57 & -10.62 & 1.69 & 1.96 & -3.54 \\
SLy4T$_{\text{min}}$       & -15.63 & -8.72 & 0.99 & -12.67 & -11.26 & -9.50 & 0.59 & -10.67 & 1.68 & 1.98 & -3.57 \\
SLy4T$_{\text{self}}$      & -15.73 & -8.59 & 1.02 & -12.67 & -11.29 & -9.50 & 0.60 & -10.70 & 1.70 & 1.97 & -3.71 \\
TZA                        & -16.57 & -7.08 & 1.36 & -12.50 & -11.33 & -9.18 & 0.72 & -10.61 & 1.90 & 1.89 & -4.65 \\
\noalign{\smallskip}\hline
\end{tabular}
\end{table*}

In Fig.~\ref{fig:ni56:tensor:decomposition2}, we decompose the tensor energy
in the cartesian and in the angular-momentum coupled representations.
Isovector contributions for this $N=Z$ nucleus are smaller than 20 keV
for all deformations. In the left panels, the contributions
corresponding to the symmetric and asymmetric terms in the cartesian
representation are plotted. Both are of the order of a few MeV for T22 and
of opposite sign to ensure a total contribution close to zero.
They are also of similar magnitude for T26, but repulsive in both cases.
Results from T24 are intermediate between those of T22 and T26.
The right panels show the contributions to the total tensor energy
from the vector and pseudotensor terms in the angular-momentum coupled
representation. The contribution from the pseudoscalar term is zero
with the symmetries assumed here. Except for T22, where the vector
contribution is zero by construction, the pseudotensor
contribution is two orders of magnitude smaller than the vector
one. We found similar results for all other parameterizations
with nonzero $C^{J1}_0$ and for all nuclei studied here. This justifies
the common practice of neglecting the pseudotensor terms for the purpose
of calculating binding energies in situations where Galilean invariance
is not an issue.

All decompositions of the tensor energy exhibit the same trend: this
energy decreases with deformation, without exhibiting much structure.
This behavior can be understood rather easily in this $N=Z$ nucleus
where the $f_{7/2^-}$ orbitals are filled at sphericity while the
$f_{5/2^-}$ ones are empty. This situation makes the tensor interaction
maximal. As soon as deformation sets in, this simple picture is
destroyed: the single-particle levels loose their purity and cross,
cf.\ the upper panel of Fig.~\ref{fig:ni56:trends}.
%
%

\subsubsection{Single-particle spectra at sphericity}

We give in Table~\ref{tab:ni56} the eigenvalues of the single-particle
Hamiltonian $\hat{h}_q$, Eq.~(\ref{eq:mf:eq}) for neutron orbitals
in the $pf$ shell obtained in calculations of \nuc{56}{Ni} imposing spherical shape.
The position of the $1g_{9/2^+}$ level is also given, although it is
far above the Fermi energy at spherical shape, as it determines the
size of the deformed gap at 28 in the Nilsson diagram through its
downsloping $j_z = 1/2$ levels, cf.\ Fig.~\ref{fig:ni56:trends}.

We also give the renormalized spin-orbit splittings
\begin{equation}
\Delta \epsilon_{\ell}
= \frac{1}{2\ell+1}
  \big( \epsilon_{j=\ell-1/2} - \epsilon_{j=\ell+1/2} \big)
\, ,
\end{equation}
which for a standard modified oscillator potential would be independent
on the quantum numbers of the single-particle states, the centroids
\begin{equation}
\epsilon^{\text{cent}}_{\ell}
=  \frac{\ell+1}{2\ell+1} \epsilon_{j=\ell+1/2}
  +\frac{\ell  }{2\ell+1} \epsilon_{j=\ell-1/2}
\, ,
\end{equation}
of spin-orbit partners for the $2p$ and $1f$ levels, the ratio
$\Delta \epsilon_{f}/\Delta \epsilon_{p}$ of the spin-orbit splittings
of the $1f$ and $2p$ levels, and the distance of their centroids
$\epsilon^{\text{cent}}_f - \epsilon^{\text{cent}}_p$.

Experimental separation energies from or into low-lying levels in the odd-$A$
neighbors of \nuc{56}{Ni} that have the characteristics of a single-particle
configuration are given in the first line of  Table~\ref{tab:ni56}.
Empirical values for the single-particle energies
are given in the second line. These quantities are usually compared to the
eigenvalues $\epsilon_i$ of the mean-field hamiltonian, although many factors
make this comparison questionable, see for instance ~\cite{Zal08a}
and Sect.~IV.~B of Article~I.
One source of ambiguity is the coupling of the
particle or hole outside the closed shell to the vibrations
of the core. Using the schematic extended unified model,
the authors of \cite{Tra96aE} attempted to remove this effect
to determine "bare" values of the
single-particle energies by reverse engineering
from the low-lying excitation spectra of \nuc{56}{Ni} and its
odd-$A$ neighbors. Although model-dependent, we
include these values here in the second line of Table~\ref{tab:ni56}
to have a rough estimate of the order of magnitude of the corresponding
corrections. In any event, for an $N=Z$ nucleus such as \nuc{56}{Ni},
there is also a contribution from the Wigner energy to the separation
energies~\cite{Cha07a}, which is not considered in Ref.~\cite{Tra96aE}.
Its main effect for a magic nucleus is to render the gap in the
separation energies much larger than the gap in the spectrum of
eigenvalues of the mean field. Using a schematic model, Chasman~\cite{Cha07a}
estimates
the correction from the Wigner energy to the size of the $N=28$ gap in the
empirical spectrum from separation energies to be larger than 2 MeV.

The results for the parameterizations T22, T26, T44 and T62 are given
in the next four lines. By construction, there is no tensor contribution
at sphericity for T22. The other three interactions share the same
isoscalar coupling constant $C^J_0 = 120$ MeV fm$^5$, but differ in
their isovector one $C^J_1$. The presence of a tensor term has a small
effect on the absolute position of the $2p$ levels, which move at most
by 200 keV, much less than the $1f$ levels for which the changes go up
to 1.2~MeV. The tensor term is mainly responsible for a reduction of
the spin-orbit splittings, whereas the centroids of the $2p$ and $1f$
levels are affected to a much smaller extent. A change in the centroid
position cannot be directly related to the tensor terms since they do not
contribute directly to the part of the mean field which governs it. The
modification of the centroids is a non-linear effect induced by the tensor.
Although small, the net effect is visible, in particular for the
distance between the $1f$ and $2p$ centroids that are pulled into
opposite directions.

The shift of the centroids is correlated to the isoscalar
tensor coupling constants (cf.\ T22 and T44), but unexpectedly for a 
$N=Z$ nucleus, also slightly to the isovector ones (cf.\ T26, T44 and T62).
The isovector densities and currents induced
by the isospin breaking Coulomb interaction are small and
do not significantly contribute to mean fields and energies. The
differences between  the centroids obtained with T26, T44 and T62
are predominantly a consequence of the readjustment of the entire
energy functional for each strength of the tensor terms.

The larger impact of the tensor terms on the $1f$ levels than on the $2p$
ones is still more apparent when a trivial angular-momentum factor
in the spin-orbit splitting $\Delta \epsilon_\ell$ is taken out.
This result has a geometrical origin discussed in Fig.~16 of
Article~I for a different example: a zero-range tensor interaction has
the largest impact on spin-orbit splittings for those levels that have the
same nodal structure as the ones that dominate the spin-current $\vec{J}$.

As discussed in Article~I and exemplified in Fig.~4, the isoscalar tensor term
has the tendency to reduce the spin-orbit splitting in spin-unsaturated nuclei
for the T$IJ$ parameterizations studied here. To maintain a given splitting,
the spin-orbit coupling constant has to be increased.
Thus, the reduction of the spin-orbit
splittings obtained with T44, as compared to those from T22, results from
the partial compensation of the change in tensor and spin-orbit
contributions. By contrast, the spin-orbit splittings obtained with
T26, T44 and T62 are fairly independent on the value of the isovector
tensor coupling constant $C^J_1$. The reason is twofold: on the
one hand, the isovector spin-current tensor density is negligibly small
in an $N=Z$ nucleus and all direct isovector contributions
to the spin-orbit field are suppressed. On the other hand, changing $C^J_1$
in the fit does not induce a significant change of the strength of the
spin-orbit interaction within the protocol of the T$IJ$ interactions.

The values obtained with SLy5 and SLy5+T are listed in the next two lines of Table~\ref{tab:ni56}.
The negative value chosen for $C^J_0$ in SLy5+T, is not compensated by a
readjustment of the spin-orbit strength and leads to a substantial increase
of all spin-orbit splittings. The negligible changes in the
position of the centroids gives an indication of the order of magnitude of
rearrangement effects from self-consistency.

\begin{figure}[t!]
\centerline{\includegraphics{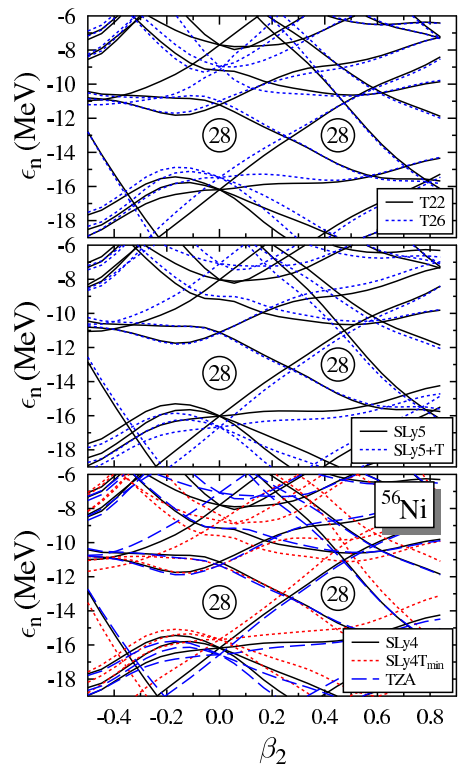}}
\caption{
\label{fig:ni56:nilsson:comp1}
(Color online)
Neutron Nilsson single-particle diagrams for \nuc{56}{Ni} 
obtained using various families of parameterizations (see text).
}
\end{figure}

The results obtained with SLy4, SLy4T, SLy4T$_{\text{min}}$,
SLy4T$_{\text{self}}$ and TZA can be found in the last five lines of
Table~\ref{tab:ni56}. Although the tensor coupling constants of
SLy4T and SLy5+T are similar, cf.\ Fig.~\ref{fig:coupling}, their
behavior with respect to the parameterizations from which
they have been constructed is quite different.
The spin-orbit splitting of the $1f$ levels obtained with SLy5+T is much
larger than with SLy5, while that of SLy4T is smaller than that of SLy4.
The good agreement with experiment obtained with SLy4T is not surprising
since \nuc{56}{Ni} is one of the data that has been used to adjust
the spin-orbit and tensor strengths. The origin of the differences between
these interactions is the additional reduction of the spin-orbit force
in SLy4T, to about $2/3$ of its original value. For \nuc{56}{Ni}, the reduced
spin-orbit interaction of SLy4T overcompensates the effect of the tensor
interaction.
The single-particle spectra obtained with SLy4T, SLy4T$_{\text{min}}$,
and SLy4T$_{\text{self}}$ differ slightly, which results from self-consistency
in the calculations and the readjustment of the other coupling constants of
the functional. Since TZA has the same tensor coupling constants as SLy4T,
but an increased spin-orbit interaction, it predicts too large spin-orbit
splittings.

The single-particle spectra from SLy4 and T22, obtained
from almost the same fit protocol, are very close as should be expected.
Also, the single-particle spectrum obtained with SLy4 lies in between
those from SLy5 and SLy5+T, as could be expected from their tensor coupling
constants, Fig.~\ref{fig:coupling}, and the similarities of the respective
fits.

Comparing calculated single-particle energies to empirical ones
from \cite{Tra96aE} and to experimental separation energies,
the fine tuning of the spin-orbit splittings
that constitutes the main difference between the interactions studied
here does not significantly improve the overall agreement with data. The main
deficiency shared by all parameterizations is that the relative
distance between the centroids of the $1f$ and $2p$ levels is nearly 2 MeV
too small, leading to different sequences of the $1f_{5/2^-}$ and $2p_{1/2^-}$
levels in calculations and data. This result is consistent with the
suspicion raised in Ref.~\cite{Ben06a} that a substantial increase of
the distance between the centroids given by SLy4
might be needed to reproduce the shape coexistence phenomena around
\nuc{74}{Kr}. However, a more careful analysis of the physics that connects
the single-particle spectra and the observable separation energies is
needed before a final conclusion can be drawn.

%
%
\subsubsection{Nilsson diagrams}

At sphericity, the single-particle spectra obtained with interactions
adjusted using the same protocol exhibit minor differences, with a
splitting of the $1f$ levels varying by about 250~keV. Variations
are slightly larger when the tensor term is added perturbatively.
The effects of these differences on the dependence of the single-particle
levels on deformation can be found in Fig.~\ref{fig:ni56:nilsson:comp1}.
Two T$IJ$ parameterizations only are plotted, as the results do not depend
on the isovector coupling constant $C^J_1$. Nilsson diagrams for protons
differ mainly by a constant shift due to the Coulomb interaction.

The most striking insight from Fig.~\ref{fig:ni56:nilsson:comp1} is
that changing the strength of the isoscalar tensor coupling modifies the
slope of the level dependence on deformation, especially close to
sphericity. The impact on deformed shell gaps depends, however,
on how a tensor term has been introduced.

For refitted parameterizations such as as T22 and T26, the difference 
between the
position of the $1f_{7/2^-}$ levels at sphericity is compensated by the
change of the slope of the single-particle levels in such a way that the
gap at $\beta_2=0.5$ has about the same size. A similar result is obtained
for all T$IJ$ and SLy$x$ parameterizations.

In contrast, for interactions with perturbatively added
or rescaled terms, the size of the deformed
gap is strongly modified: it becomes smaller for SLy5+T, and larger for
SLy4T$_{\text{min}}$. The origin of the latter difference is the reduced
spin-orbit interaction for SLy4T$_{\text{min}}$. With the same tensor
coupling constants as SLy4T$_{\text{min}}$, TZA leads to results similar
to those of the T$IJ$ parameterizations.

Similar results and similar observations can be made are obtained 
for all other nuclei discussed hereafter.

\begin{figure}[t!]
\centerline{\includegraphics{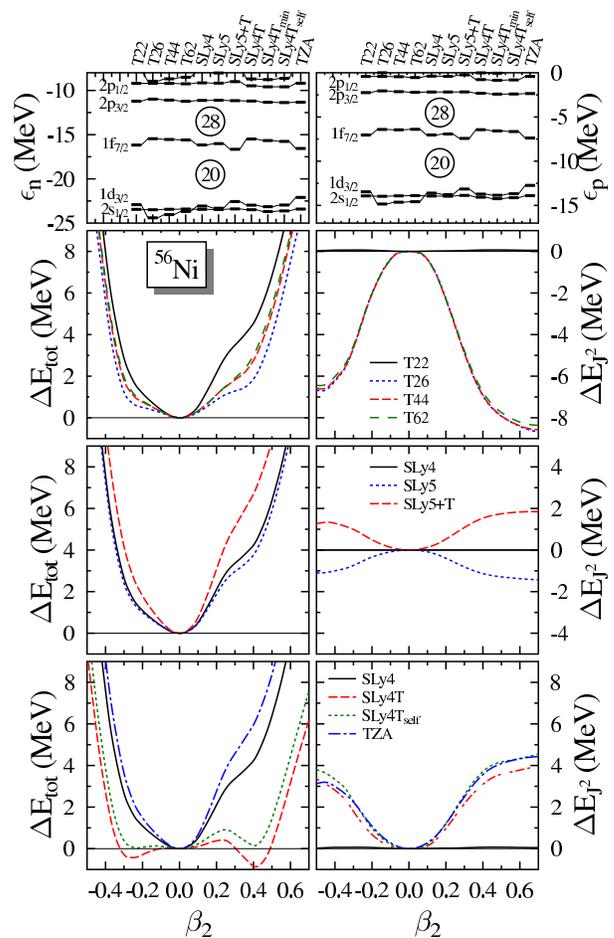}}
\caption{
\label{fig:ni56:surface:comp1}
(Color online)
The single-particle spectra at spherical shape are shown in the upper panel
for neutrons (left) and protons (right) The deformation energy (left)
and the variation of the total tensor energy (right) are plotted on the lower panels
for \nuc{56}{Ni} and different Skyrme parameterizations, as indicated.
}
\end{figure}

%
%
\subsubsection{Deformation energy and its tensor term contribution}

Deformation energy curves  are plotted in Fig.~\ref{fig:ni56:surface:comp1}.
To facilitate their correlation with shell structure, the corresponding
neutron and proton single-particle spectra at sphericity are
given in the upper panels. The total deformation energy is given on
the left-hand-side in the lower panels, whereas the difference
between the tensor energy contribution at a given deformation
and at sphericity is provided on the right-hand-side.
This tensor energy contribution decreases with deformation for T$IJ$
parameterizations and for SLy5, for which $C^J_0$  is positive. It
increases for the other parameterizations that have a negative $C^J_0$.

The differences between the energy curves are not directly linked
to the evolution of the tensor energy. They can, in fact, be related to the
shell effects that are seen in Fig.~\ref{fig:ni56:nilsson:comp1} and,
in particular, to the relative size of the spherical and prolate deformed
$N=Z=28$ gaps. A well-defined spherical minimum and a pronounced shoulder
at a deformation around $\beta_2=0.5$ are obtained for SLy4, SLy5 and
the T$IJ$ parameterizations. This structure in the energy curve is at a
position qualitatively in agreement with the properties of a
superdeformed rotational band observed in \nuc{56}{Ni}~\cite{Ru99a}.

The combined effect of the reduced spherical $N=Z=28$ gap and the large
deformed gap obtained with SLy4T has the unphysical consequence that
the ground state corresponds to a superdeformed minimum.
This result is related to a deficiency of SLy4 on which
the tensor interaction has no effect. This interaction, indeed,
predicts too small a distance between the centroids of the
$1f$, $2p$ orbitals (see Table~\ref{tab:ni56}).
In such a case, reproducing
the empirical data for the splitting of the $1f$ levels does not
guarantee a realistic shell structure. In practice, the gap between the $1f_{7/2^-}$
and $2p_{3/2^-}$ levels becomes too small, whereas the distance between
the $1f_{7/2^-}$ and $1g_{9/2^+}$ levels is now too large.

The parameterization SLy5+T has the inverse drawback: the gap at $28$
is too large at sphericity, preventing the formation of a secondary
gap at large deformation.

The deformation energy and the relative change of the tensor terms obtained
with SLy4T (shown) and with SLy4T$_{\text{min}}$ (not shown) cannot be
distinguished within the resolution of Fig.~\ref{fig:ni56:surface:comp1}.
This is less obvious than one might think. The refit that leads from SLy4T
to SLy4T$_{\text{min}}$ changes the absolute binding energy of \nuc{56}{Ni}
by nearly 6 MeV from $-469.522$ (SLy4T) to $-475.480$ MeV
(SLy4T$_{\text{min}}$).

The results obtained with  SLy4T$_{\text{self}}$ and TZA confirm the crucial
role of the spin-orbit strength. Both interactions have been adjusted with
the same protocol and the SLy4T
values for the tensor coefficients, but  SLy4T$_{\text{self}}$  shares the same
spin-orbit strength as SLy4T while it has been freely varied for TZA. This
variation of the spin-orbit leads to results quite close to those of the
T$IJ$ parameterizations for TZA, in contrast to what is obtained with
SLy4T$_{\text{self}}$.

\begin{table}[t!]
\caption{
\label{tab:coup:choice}
Tensor coupling constants of T44 and three extensions of T44
beyond the spherical symmetry constructed using the freedom
of choice given by an energy functional (see text).
All values are in MeV fm$^{5}$. Coupling constants are given
for all three representations of the tensor part of the energy
functional, Eqns.~(\ref{eq:EF:tensor1}) and~(\ref{eq:func:Skyrme:pn}).
All coupling constants not shown are identical.
}
\begin{tabular}{lcrrr}
\hline\noalign{\smallskip}
           & T44  &  T44 II & T44 III & T44 IV \\
\noalign{\smallskip}\hline\noalign{\smallskip}
$C^{J0}_0$ &  -20.994 &  -60 &   40 & -160 \\
$C^{J0}_1$ &   50.027 &    0 &    0 &    0 \\
$C^{J1}_0$ &   60     &   60 &   60 &   60 \\
$C^{J1}_1$ &    0     &    0 &    0 &    0 \\
$C^{J2}_0$ &   46.806 &    0 &  120 & -120 \\
$C^{J2}_1$ &   62.433 &    0 &    0 &    0 \\
\noalign{\smallskip}\hline\noalign{\smallskip}
$C^T_0$    &  -83.403 &  -60 & -120 &    0 \\
$C^T_1$    &  -31.216 &    0 &    0 &    0 \\
$C^F_0$    &   73.194 &  120 &    0 &  240 \\
$C^F_1$    &  -62.433 &    0 &    0 &    0 \\
\noalign{\smallskip}\hline\noalign{\smallskip}
$b_{14}$   &   52.187 &   60 &  120 &    0 \\
$b_{15}$   &   62.433 &    0 &    0 &    0 \\
$b_{16}$   &  -67.813 &  -60 &    0 & -120 \\
$b_{17}$   &   62.433 &    0 &    0 &    0 \\
\noalign{\smallskip}\hline
\end{tabular}
\end{table}

%
%
\subsubsection{The freedom of using an energy density functional}

As discussed in Sect.~\ref{sect:param}, the fits of tensor couplings
have all been performed
assuming spherical symmetry. Two coupling constants have been fixed in this
way, either $t_e$ and $t_o$ or $C^{J1}_t$, $t=0$, 1. There remains the
freedom to choose the $C^{J2}_t$ coefficients. It has been assumed, for 
the interactions
$TIJ$, that there are underlying two-body central, spin-orbit and tensor
forces, cf.\ Sect.~\ref{sect:choices}. In this case, there is a one-to-one
correspondence between the $C^{J1}_t$ coefficients and parameters $t_1$, 
$x_1$, $t_2$,
$x_2$ of the central Skyrme force and parameters $t_e$ and $t_o$ of the
tensor force. With this choice, all coupling constants of the energy
functional, $C^{J2}_t$, $t=0$, 1, or, alternatively, the $C^T_t$ and 
$C^F_t$ coupling are also fixed. As explained in Sect.~\ref{sect:choices},
other choices can be made.
To explore the impact of doing so, we have constructed three variants
of the parameterization T44, listed in Table~\ref{tab:coup:choice}.
These consist of setting either $C^{J2}_0$ (case II), or $C^F_0$
(case III), or $C^T_0$ to zero (case IV). Each choice leads to
different values for the coupling constants of the pseudoscalar and
pseudotensor part of the tensor terms. We concentrate here on the
isoscalar part of the functional, since for the purpose of our study
isovector effects are negligible in a $N=Z$ nucleus.

\begin{figure}[t!]
\centerline{\includegraphics{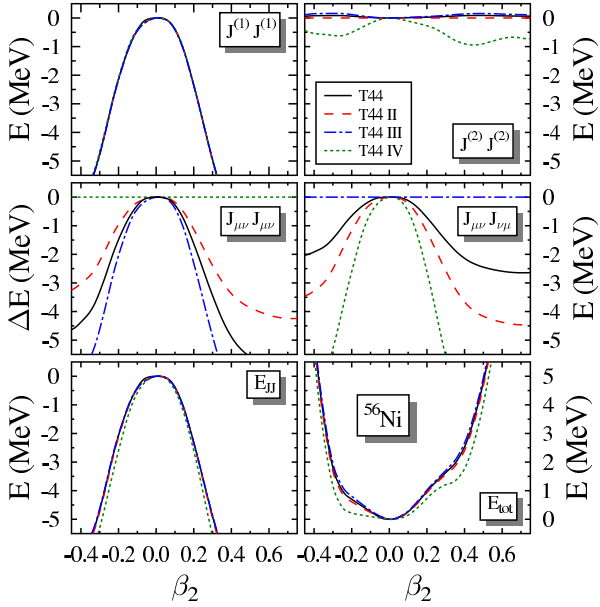}}
\caption{
\label{fig:coup:choice:ni56}
(Color online)
Deformation energy curve (lower right), tensor  energy
(lower left), and its decomposition into symmetric, asymmetric, vector and
pseudotensor parts  for \nuc{56}{Ni} as obtained with the variants
of T44 defined in Table~\ref{tab:coup:choice}.
}
\end{figure}

The total deformation energy and the variation of tensor contributions
with respect to their value at sphericity are plotted
in Fig.~\ref{fig:coup:choice:ni56} as a function of deformation. The
total tensor energies (bottom left), and the decomposition into vector
and pseudotensor contributions on the one hand (top), and into symmetric
and asymmetric cartesian components on the other hand (middle) are 
presented.

As expected from the smallness of the pseudotensor contribution in
Fig.~\ref{fig:ni56:tensor:decomposition2}, the choice II where the coefficient
$C^{J2}_0$ is set to zero does not lead to any sizable difference with T44.
The results for the two other choices are less obvious. As can be seen on
the middle panels of Fig.~\ref{fig:coup:choice:ni56}, the cartesian
components of the tensor energy are both sizable and one could expect that
setting one of the two coefficients of these terms to zero will have a large
effect. One can see in the middle panels of Fig.~\ref{fig:coup:choice:ni56}
that both terms show significant variation with respect to deformation. The
decomposition is even significantly different for T44 and choice II, although
leading in both cases to the same total energy. Self-consistency effects are
such that for the choices III (where $J_{q,\mu\nu}J_{q,\nu\mu}$ has a
coefficient equal to zero) and IV  (where $J_{q,\mu\nu}J_{q,\mu\nu}$ has
a coefficient equal to zero), the remaining contribution has a variation
very similar to the variation of the total tensor energy of T44.

The way a parameterization is extended beyond sphericity has a small, but
visible, effect on the variation of the total energy with deformation.
Setting the coefficient of $J_{q,\mu\nu} J_{q,\mu\nu}$ equal to zero,
(option IV) significantly softens the energy curve with an oblate shoulder
nearly degenerate with the spherical configuration. The extension of a
parameterization beyond sphericity is not a trivial choice and the procedure followed
for such extension should always be made transparent. The small
differences seen in a simple nucleus such as \nuc{56}{Ni} could become more
dramatic in other nuclei.

%
%
\subsection{Doubly-magic nuclei}
%
%
\subsubsection{\nuc{40}{Ca}}

\begin{figure}[t!]
\centerline{\includegraphics{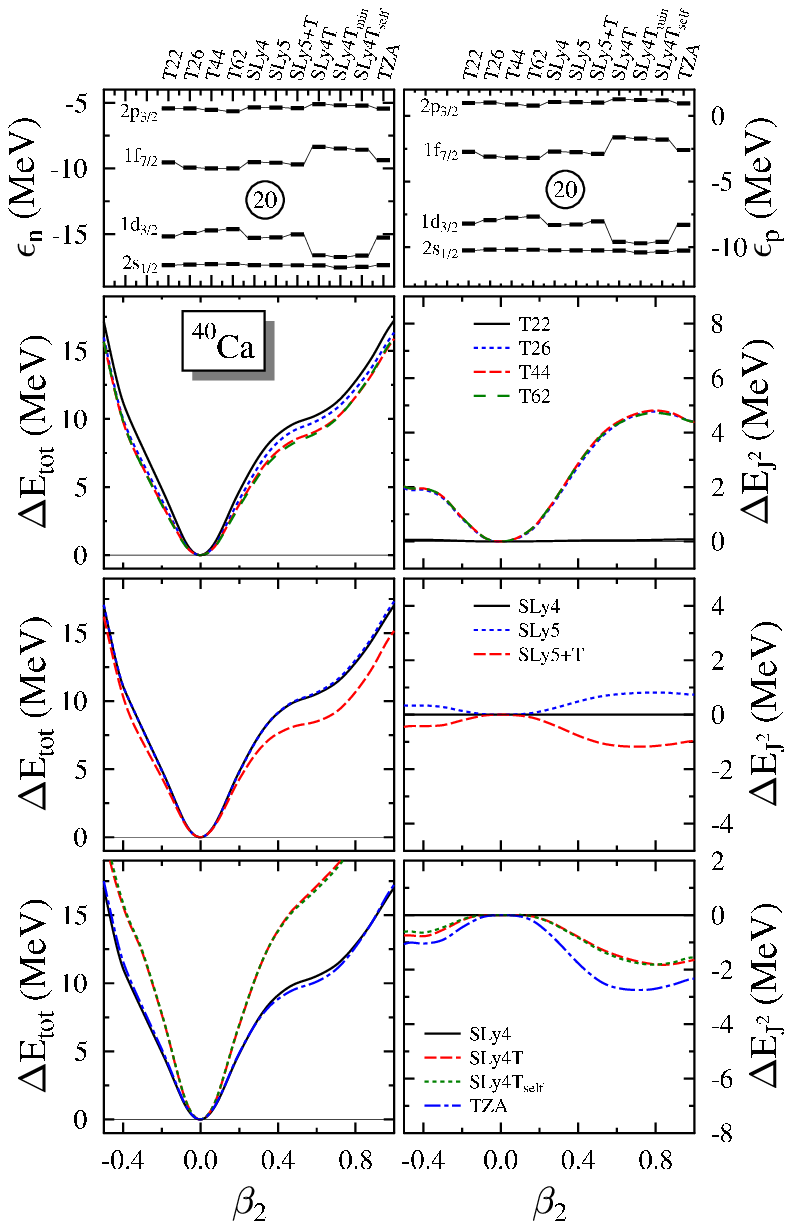}}
\caption{
\label{fig:ca40:surface}
(Color online)
Same caption as Fig.~\ref{fig:ni56:surface:comp1}, but for \nuc{40}{Ca}.
Note that the scale of the total deformation energy differs from the one of the
tensor terms  by a factor two.
}
\end{figure}

The $N=Z=20$ nucleus \nuc{40}{Ca} is the heaviest known doubly-magic
nucleus which exhibits oscillator shell closures. The variation
of the total  energy, the change of the contribution to the energy of
the tensor terms, and the single-particle spectra at spherical shape are
plotted in Fig.~\ref{fig:ca40:surface}.
The configuration of \nuc{40}{Ca} is spin-saturated at sphericity,
and the corresponding tensor energy is only due to small
effects as pairing and non-identity of the single-particle
wave functions of spin-orbit partners. Spin saturation disappears as soon
as the nucleus is deformed, but the filling of single-particle
states remains identical for protons and neutrons.
Thus, the contribution of the tensor terms to the
total binding energy induced by deformation is almost purely isoscalar
and has the same sign as the isoscalar coupling constant $C^J_0$.
The tensor term is close to zero at sphericity,
and increases with deformation. As the coupling constant
$C^{J1}_0$ of the isoscalar vector part of the tensor terms
is the same for T26, T44 and T62,
$C^J_0 = 120$ MeV fm$^5$, these interactions give
nearly identical tensor contributions to the total energy.

\begin{figure}[t!]
\centerline{\includegraphics{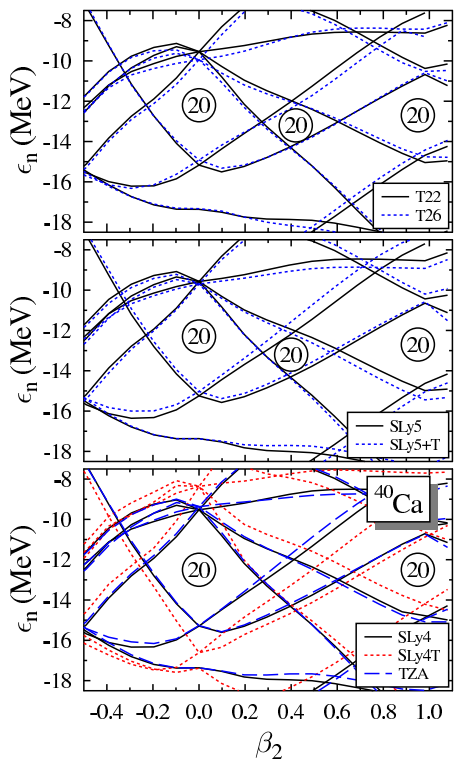}}
\caption{
\label{fig:ca40:nilsson:comp1}
(Color online)
Same as Fig.~\ref{fig:ni56:nilsson:comp1}, but for \nuc{40}{Ca}
}
\end{figure}

Despite this feature, the energy curves corresponding to T26, T44 and
T62, are not identical. One can see that the softness of the energy curves
increases with the strength of the isovector part of the tensor interaction.
Once again, these changes result from the readjustment of the coupling
constants of all terms of the energy functional when varying $C^J_0$
and $C^J_1$ to a sample of data that includes both $N=Z$ and $N \neq Z$
nuclei. In fact, as for \nuc{56}{Ni} discussed in
Figs.~\ref{fig:ni56:total:decomposition}
and~\ref{fig:ni56:total:decomposition2}, most components of the energy
show larger variations between the parameterizations than the tensor
energy. This example illustrates particularly well the fact that the impact of
the tensor terms on deformation energies cannot be foreseen from the
sole knowledge of the tensor coupling constants and of the variation
of the degree of spin saturation with deformation. The energy of the
shoulder that can be seen  at a $\beta_2$ value around 0.5 is
significantly lowered for the tensor interactions T44 and T62. This
shoulder can be related to the existence of a superdeformed band in
\nuc{40}{Ca}~\cite{Ide01aE,Chi03aE}.

The isoscalar tensor coupling  is attractive for SLy5+T
and SLy4T. The energy curve obtained with SLy5+T is only slightly
different from the one calculated with SLy5 around sphericity, but presents
a significant lowering of the energy of the shoulder. The situation is
quite different for SLy4T. The energy curve varies in opposite direction
from the tensor energy and is stiffer than that obtained with SLy4. In
particular, the shoulder is obtained at a much larger excitation energy.
This clearly is a consequence of the reduced spin-orbit
interaction, as this feature of SLy4T is shared by SLy4T$_{\text{self}}$,
but not TZA. The energy of the tensor term is
larger for the parameterization TZA than for SLy4T, although the
tensor coefficients  have the same values in both cases. The increase
(in absolute value) of the tensor energy is compensated by
changes in other terms of the energy functional, in particular
a slight reduction of the spin-orbit term, in such a way that the
energy curves obtained with SLy4 and TZA are nearly undistinguishable.

The qualitatively different deformation dependence of the tensor
energy found for \nuc{56}{Ni} and \nuc{40}{Ca} is accompanied by
systematic differences in the single-particle levels
given in Fig.~\ref{fig:ca40:nilsson:comp1}.
As \nuc{40}{Ca} is spin-saturated, the spectrum at sphericity is nearly
the same for all parameterizations, except SLy4T. The contributions of
the tensor terms are indeed small, although not exactly zero. 
The largest differences between the results obtained with the T$IJ$ 
parameterizations are those for T22 and T26.  They are related to 
the slight difference between their coupling constants and,
in particular, to the larger strength of the spin-orbit of T26
compared to T22, cf.\ Article~I.
This effect of the spin-orbit interaction on the single-particle levels
is more drastic when its strength is explicitly adjusted to
spin-orbit splittings in this mass region, as illustrated by
the comparison between SLy4T and SLy4.

As for \nuc{56}{Ni}, Fig.~\ref{fig:ni56:nilsson:comp1}, the tensor
interaction affects the slope of the single-particle levels shown
in Fig.~\ref{fig:ca40:nilsson:comp1} as a function of deformation.
However, the changes in slopes for a given parameterization are opposite
for both nuclei. This is related to the difference in the tensor
contribution to the spin-orbit field: it increases with deformation
for \nuc{40}{Ca}, while it decreases in \nuc{56}{Ni}. The change in
slope at least partly compensates the differences found at spherical
shape when going to deformed ones. As a consequence, the spectra
around the Fermi energy at large deformation are close for refitted
parameterizations such as T22, T26 and TZA. On the contrary, the
differences between the spectra  for a perturbative interaction
such as SLy5+T and the original one increases with deformation.
Differences are larger at all deformations for SLy4T compared to
SLy4 and TZA, as a consequence of a perturbative modification of
the tensor and the spin-orbit parameterizations.

\begin{figure}[t!]
\centerline{\includegraphics{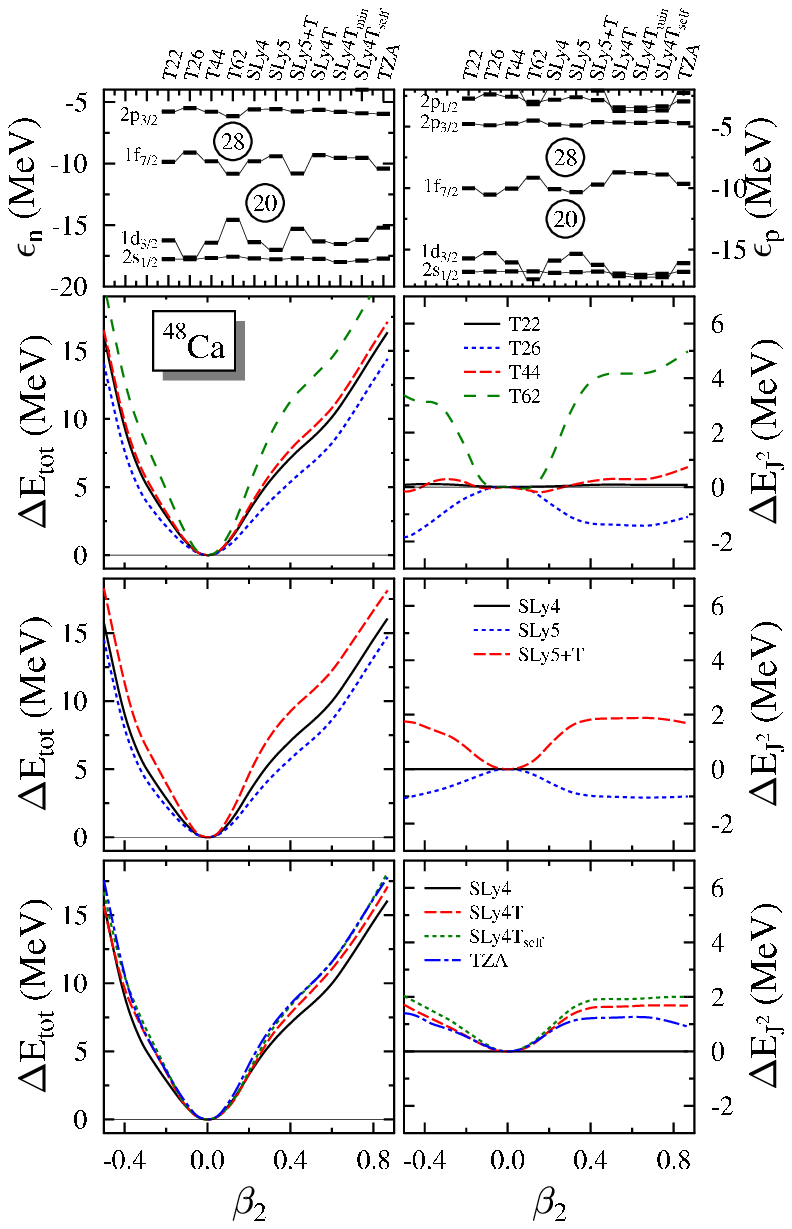}}
\caption{
\label{fig:ca48:surface}
(Color online)
Same caption as Fig.~\ref{fig:ni56:surface:comp1}, for \nuc{48}{Ca}.
Note that the scale of the total deformation energy differs from the one of the
tensor terms  by a factor two.
}
\end{figure}

\begin{figure}[t!]
\centerline{\includegraphics{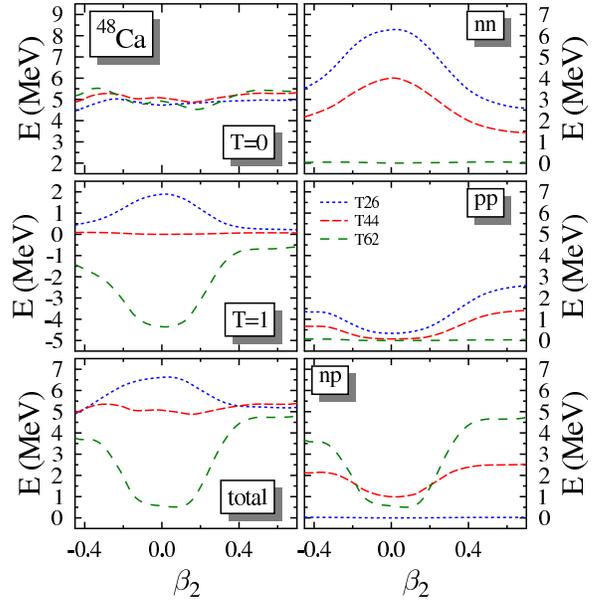}}
\caption{
\label{fig:ca48:tens:decomposition}
(Color online)
Decomposition of the total contribution of the tensor terms
to the total energy (lower left)
into its isoscalar ($T=0$) and isovector ($T=1$) parts
(upper and middle left), and into its nn, pp and np contributions
(right panels) for \nuc{48}{Ca} and for the parameterizations
T26, T44 and T62.
}
\end{figure}

%
%
\subsubsection{\nuc{48}{Ca}}

The $Z=20$, $N=28$ nucleus \nuc{48}{Ca} is spin saturated in
protons and unsaturated in neutrons at sphericity. The variations
of the tensor energy and of the total energy with deformation are
given in Fig.~\ref{fig:ca48:surface}. The upper panels show that the
four T$IJ$ parameterizations behave very differently
in contrast to the case of \nuc{40}{Ca}. The tensor energy
is nearly independent of deformation for T44, whereas
it decreases with deformation for T26 and increases for T62. As confirmed
by the behaviors of the $T=0$ and $T=1$ components of the tensor energy
that are plotted in Fig.~\ref{fig:ca48:tens:decomposition}, the isoscalar
contribution to the tensor energy does not vary much for all T$IJ$
parameterizations, whereas the isovector contribution presents an 
extremum at sphericity and goes rapidly to zero with deformation.
One can relate the behavior of the tensor energy
to the fact that the spin-current density is the largest for neutrons
at sphericity, but nearly zero for protons. With increasing deformation
the proton  spin-current density grows, whereas the neutron one is
reduced, as can be deduced from the decomposition of the tensor
terms into their nn, pp and np contributions provided in
Fig.~\ref{fig:ca48:tens:decomposition}. Note for T62 that the nn and
pp contributions are close to zero at all deformations by construction,
but this does not result in the proportionality of $T=0$ and $T=1$
components with respect to each other and to the np contribution. 
In general, the
nn and pp contributions to both $T=0$ and $T=1$ part are nonzero and
deformation dependent.
Similarly, from the (by construction) nearly vanishing np contribution
found at all deformations for T26 it also cannot be concluded that
the $T=0$ and $T=1$ components are proportional to each other and the
sum of the nn and pp contributions.

For \nuc{48}{Ca}, the differences between the deformation energy curves
in Fig.~\ref{fig:ca48:surface} are clearly correlated with
the variations in the isovector tensor energy.
The curve is softer for a repulsive isovector contribution to the
tensor energy, as it is for T26, and stiffer for an attractive
isovector contribution, as it is for T62. This counterintuitive
outcome is the consequence of the rapid decrease of the isovector
tensor energy from large values at sphericity to very small values
with deformation, cf.\ Fig.~\ref{fig:ca48:tens:decomposition}.
The results obtained with the interactions SLy4T and SLy5+T
are plotted in the lower panels of Fig.~\ref{fig:ca48:surface}.
The isovector tensor term has the same sign for these two interactions
as for T62 leading also to stiffer energy curves. However, the magnitude
of the effect is smaller, the isovector coupling constants being smaller.

Compared to all other interactions, but T62, the SLy4T, SLy4T$_{\text{min}}$
and SLy4T$_{\text{self}}$ functionals give a larger $Z=20$ gap at the
expense of a reduced $Z=28$ one. This is the consequence of the tightly
adjusted spin-orbit splittings of the $1f$ levels in \nuc{40}{Ca},
\nuc{48}{Ca}, and \nuc{56}{Ni} through an attractive tensor interaction
in conjunction with a reduced spin-orbit
force. The reduced spin-orbit interaction also switches the ordering of the
$1d_{3/2^+}$ and $2s_{1/2^+}$ levels below the $Z=20$ gap, at variance with
empirical data. Keeping the negative tensor coupling constants
of SLy4T, but allowing for the readjustment of the spin-orbit strength in TZA
brings the level spacings back to values close to the ones of the original SLy4
interaction.

%
%
\subsubsection{\nuc{68}{Ni}}

It is usually assumed from its spectrum that \nuc{68}{Ni}
is doubly-magic~\cite{Bro95aE,Ish00aE,Sor02aE}.
The excitation energy of its first $2^+$ state is, indeed, large and the
$B(E2; 0^+_{\text{gs}} \rightarrow 2^+_1)$ value small~\cite{Bre08aE}.
However, while the $Z=28$ proton shell closure is clearly visible in the
mass systematics along the $N=40$ isotonic chain~\cite{Rah07aE}, there
is no pronounced discontinuity in the masses of Ni isotopes when
crossing $N=40$~\cite{Rah07aE,Gue07aE}, which hints at a more complex
situation. An alternative explanation of the properties of the first
$2^+_1$ state in \nuc{68}{Ni} is based on the impossibility to construct
the first $2^+$ state as a simple neutron 1p-1h excitation. Indeed, the
odd-parity $pf$ shell is completely filled and the $1g_{9/2^+}$ orbital 
empty~\cite{Gra01aE,Gra01bE} and at least a 2p-2h excitation is needed to
construct a positive-parity state. The interpretation of \nuc{68}{Ni}
as a doubly-magic nucleus has also been questioned by shell model and
QRPA calculations~\cite{Lan03a}. Let us mention, finally, that the first
excited state of \nuc{68}{Ni} is a $0^+$ level with a small $B(E0)$ value
to the ground state \cite{Kib05aE}, pointing to a possible shape
coexistence with weak mixing.

\begin{figure}[t!]
\centerline{\includegraphics{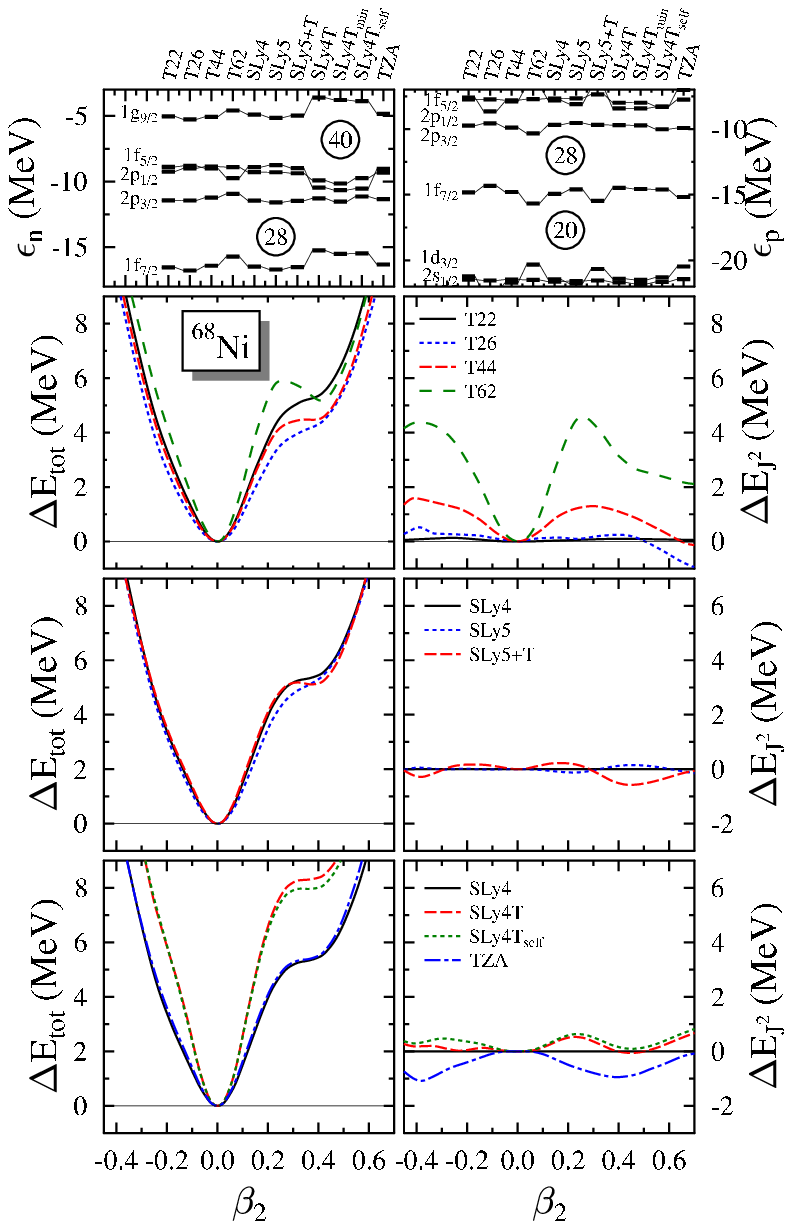}}
\caption{
\label{fig:ni68:surface}
(Color online)
Same caption as Fig.~\ref{fig:ni56:surface:comp1}, but for \nuc{68}{Ni}.
}
\end{figure}

The deformation energy curves and the variation of the tensor energy with
deformation are plotted in Fig.~\ref{fig:ni68:surface} for \nuc{68}{Ni}.
With $N=40$ and $Z=28$, this nucleus is spin-saturated for neutrons and
spin-unsaturated for protons.

\begin{figure}[t!]
\centerline{\includegraphics{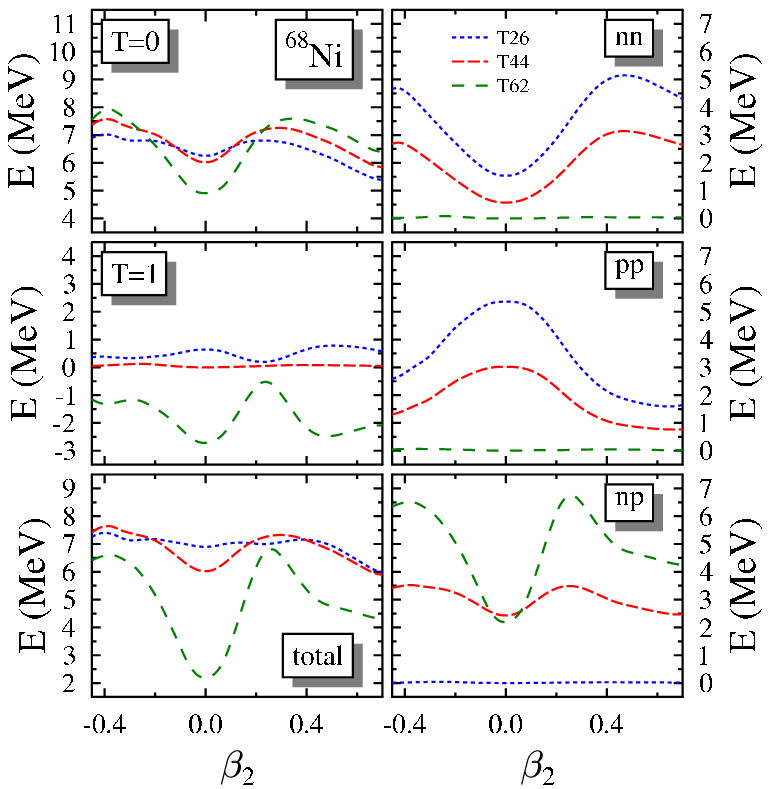}}
\caption{
\label{fig:ni68:tens:decomposition}
(Color online)
Same caption as Fig.~\ref{fig:ca48:tens:decomposition}, but for \nuc{68}{Ni}.
}
\end{figure}

It is instructive to compare \nuc{68}{Ni} and \nuc{48}{Ca}
(Fig.~\ref{fig:ca48:surface}). Both nuclei have, indeed, similar
single-particle configurations. The $1f_{7/2^-}$ subshell is completely
filled for protons in \nuc{68}{Ni} and for neutrons in \nuc{48}{Ca},
and both nuclei are spin-saturated for the other type of nucleons.
However, the comparison of both nuclei in fact indicates large
differences. The variance can be related to two factors. First,
the degeneracy of the shells that makes the $N=40$ gap for neutrons is
much larger than those that make the $Z=20$ gap for protons. Second,
the $Z=20$ and $N=40$ gaps have different sizes. The latter is not
large enough to suppress pairing correlations, such that the spin-saturation
is broken and the neutron spin-current density is non-negligible
at spherical shape, in particular for the T26 parameterization 
predicting the smallest $N=40$
gap. The main consequence is that for \nuc{68}{Ni} all contributions
to the tensor energy at spherical shape are nonzero, unless suppressed
by their coupling constant, see  Fig.~\ref{fig:ni68:tens:decomposition}.
Also, the np component
is not just increasing with deformation but fluctuating, most obviously
for T62, where it is the only sizable non-zero contribution.
By contrast, the rapidly varying nn and pp contributions for T26
fortuitously add up such that this parameterization presents the
smallest variation of the tensor energy among the parameterizations
with the same isoscalar coupling constant. The behavior of the tensor
energy is reflected in the total energy curves in Fig.~\ref{fig:ni68:surface}:
it creates an inflexion of the energy curve at $\beta_2 \approx 0.4$,
which is sufficiently large for T62 to create a secondary minimum.

The  tensor energy varies less with deformation for both the SLy5 
and SLy5+T interactions;
therefore, the energy curves obtained with both of these parameterizations
and with SLy4 are very similar. The gap at $N=40$ obtained with
SLy4T is larger, having the size of a major shell closure.
It results in a very stiff energy curve. Such a behavior
is not directly related to the variation of the tensor
energy with deformation, but to the small spin-orbit strength,
as shown by comparing the results of SLy4T$_{\text{self}}$
and TZA, just like for \nuc{40}{Ca}.
%
%
\subsubsection{\nuc{78}{Ni}}

\begin{figure}[t!]
\centerline{\includegraphics{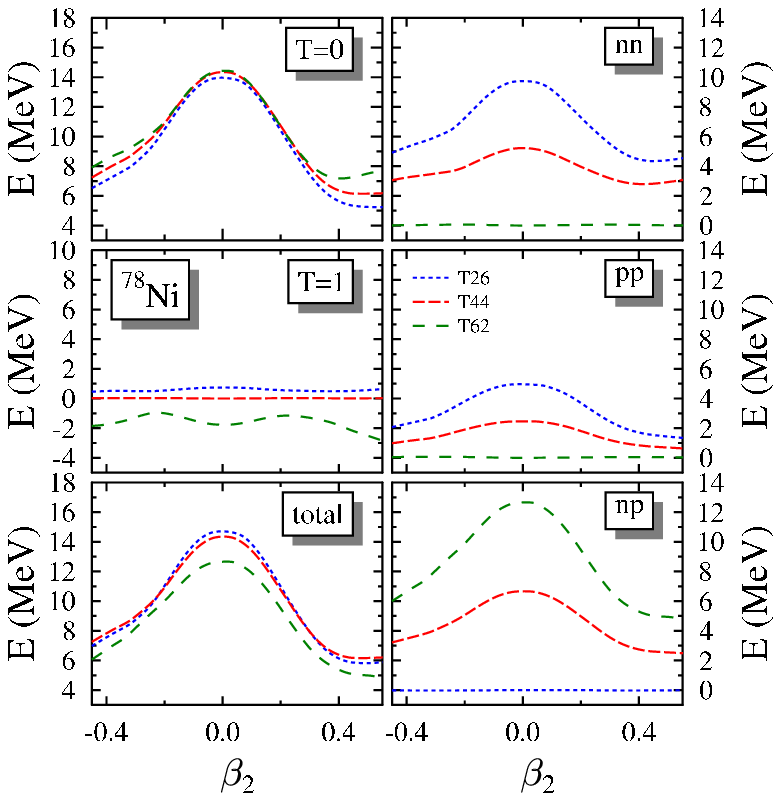}}
\caption{
\label{fig:ni78:tens:decomposition}
(Color online)
Same caption as Fig.~\ref{fig:ca48:tens:decomposition}, but for \nuc{78}{Ni}.
}
\end{figure}

\begin{figure}[t!]
\centerline{\includegraphics{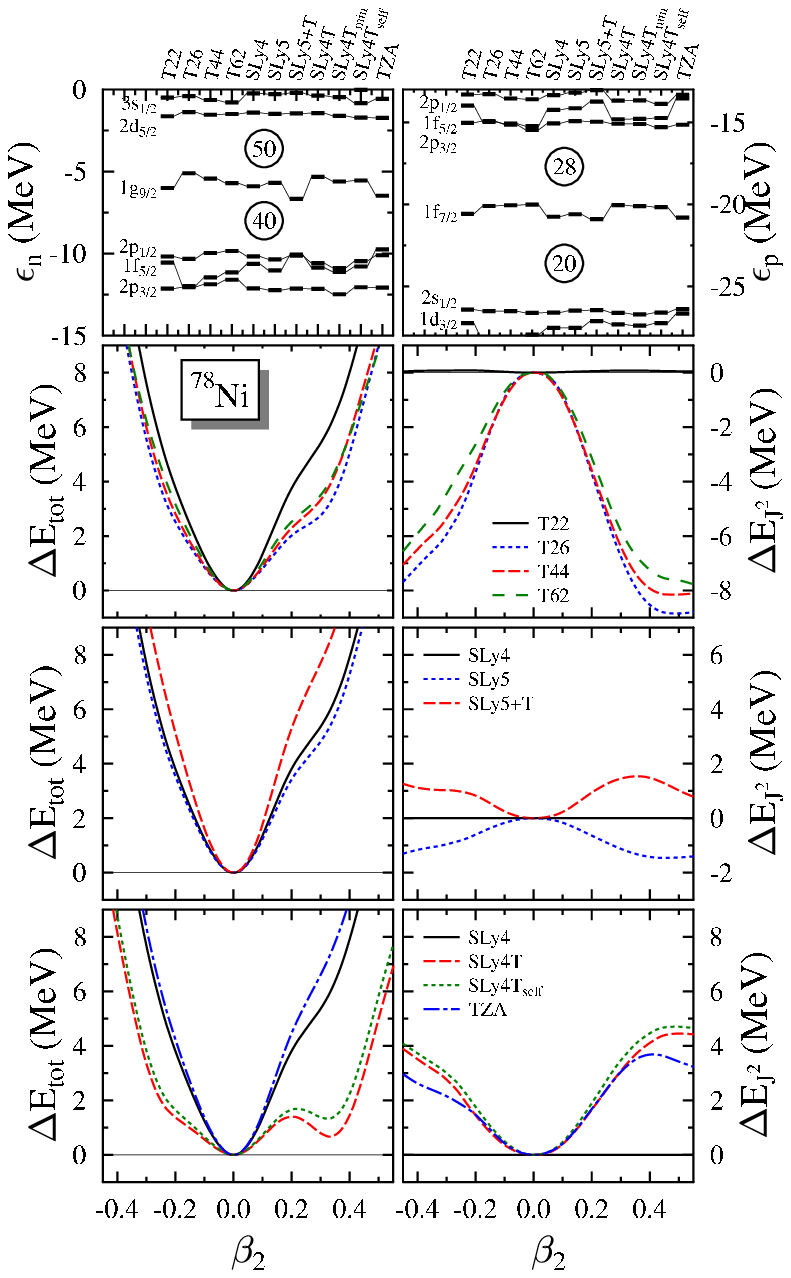}}
\caption{
\label{fig:ni78:surface}
(Color online)
Same caption as Fig.~\ref{fig:ni56:surface:comp1}, for \nuc{78}{Ni}.
}
\end{figure}

Although it has been observed for the first time more than a decade
ago~\cite{Eng95aE}, not much is yet known about the neutron-rich \nuc{78}{Ni}
besides its existence and its $\beta$-decay half-life~\cite{Hos05aE}.
The relatively long half-life and the systematics of the separation energies
to and from $N=50$ isotones down to $Z=30$ suggests that the $N=50$
shell closure persists for \nuc{78}{Ni}~\cite{Hak08a}. In such a case,
\nuc{78}{Ni} is spin non-saturated both in protons at $Z=28$ and neutrons
at $N=50$. The spin-currents from the unsaturated proton $1g_{9/2^+}$
and neutron $1f_{7/2^-}$ levels point into the same direction. Their
degeneracies differ by only two, and their nodeless radial wave functions
are sufficiently similar that their contributions to the isovector
spin-current density nearly cancel. This suggests that the isovector
tensor terms cannot play a decisive role in \nuc{78}{Ni}, in
spite of this nucleus' large isospin asymmetry. The isovector
decomposition of the tensor terms for T26, T44 and T62 shown in
Fig.~\ref{fig:ni78:tens:decomposition} indeed confirms that the
$T=1$ contribution is small for all deformations. The $T=0$ contribution
dominates and is nearly the same for all three parameterizations,
such that the different relative weight of the nn, np and pp
contributions does not play a significant role. One can,
therefore, expect that its energy curves depend on
the tensor parameterization in a way similar to \nuc{56}{Ni}.
Figure~\ref{fig:ni78:surface} indicates that this is, indeed, the case for
all the interactions that we have studied. In particular, all
interactions predict the behavior of a doubly-magic
nucleus. The T$IJ$ parameterizations with non-zero $C^J_0$ values lead
to softer energy curves than T22, SLy4 and SLy5, with an inflexion
point around $\beta_2=0.3$, whereas SLy5+T gives a much stiffer
deformation energy curve.
The difference between the total deformation energy curves from
T26, T44 and T62 is even smaller for \nuc{78}{Ni} than what was found
for \nuc{56}{Ni}. This suggests that the readjustment and self-consistency
effects at its origin for \nuc{56}{Ni} are compensated by the asymmetry
in \nuc{78}{Ni} just in  such a way that the net isospin dependence 
vanishes for this nucleus.

The parameterizations SLy4T and SLy4T$_{\text{self}}$ give again
results that are qualitatively different from those of the others:
they lead to a very pronounced deformed minimum at an excitation energy
around 1~MeV, although the spherical gaps at $N=50$ and $Z=28$ are
not smaller than those from T26, for example. This is again a consequence
of the reduced contribution of the spin-orbit interaction to the
deformation energy for these interactions.

\begin{figure}[t!]
\centerline{\includegraphics{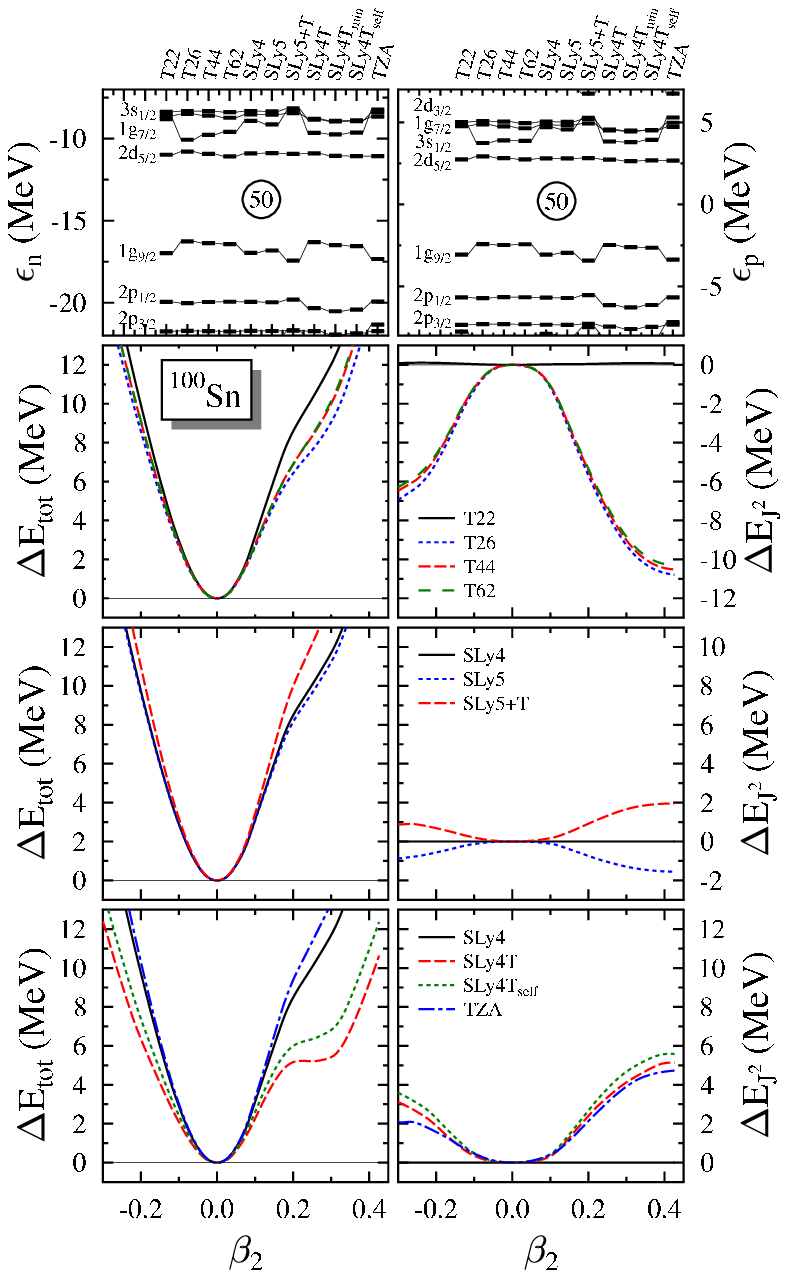}}
\caption{
\label{fig:sn100:surface}
(Color online)
Same caption as Fig.~\ref{fig:ni56:surface:comp1}, but for \nuc{100}{Sn}.
}
\end{figure}

%
%
\subsubsection{\nuc{100}{Sn}}

The proton-rich and probably heaviest bound $N=Z$ doubly-magic nucleus
\nuc{100}{Sn} has been observed more than a decade
ago~\cite{Schn94aE,Lew94aE}. Up to now, the only spectroscopic information
in the direct vicinity of the nucleus is a 172 keV $\gamma$ ray observed
in \nuc{101}{Sn} \cite{Sew07aE}.
It has been tentatively interpreted as corresponding to the transition 
between an excited $7/2^+$ level to the $5/2^+$ ground state
although the order of these two levels is not firmly established.
In any case, the distance between these two levels is much lower than the
energy difference between the spherical $1g_{7/2^+}$ and
$2d_{5/2^+}$ orbitals predicted by all Skyrme parameterizations plotted in
Fig.~\ref{fig:sn100:surface}. The tensor interaction has some effect on this
spacing: it decreases from more than 1~MeV for T22 down to 600 keV for T26.
Of course, the comparison between the energy levels in \nuc{101}{Sn} and
the single-particle energies supposes that both are pure single-particle
configurations which is far from being established. Similar
discrepancies with other parameterizations of the self-consistent
mean field were reported in Ref.~\cite{Sew07aE}. From this, however,
one cannot safely draw the conclusion that T26 is the most realistic
among the parameterizations studied here. The distance between the
$1g_{7/2^+}$ and $2d_{5/2^+}$ levels depends on the balance between
the distance of the centroids of the $1g$ and $2d$ levels as well as
on their respective spin-orbit splittings, none of which can be expected
to be described well throughout the chart of nuclei for any of the
current parameterizations of the Skyrme EDF, see Article~I
and~\cite{Kor08a}.

The variation of the deformation energy and of the energy contribution
of the tensor terms with quadrupole deformation for \nuc{100}{Sn} are
presented in Fig.~\ref{fig:sn100:surface}. For all parameterizations, the
results are very similar to those obtained for \nuc{56}{Ni},
Fig.~\ref{fig:ni56:surface:comp1}, and \nuc{78}{Ni},
Fig.~\ref{fig:ni78:surface}. The main difference
is that the structure appearing at moderate deformation
in the total deformation energy surface is less pronounced and 
located at higher excitation energies.

\begin{figure}[t!]
\centerline{\includegraphics{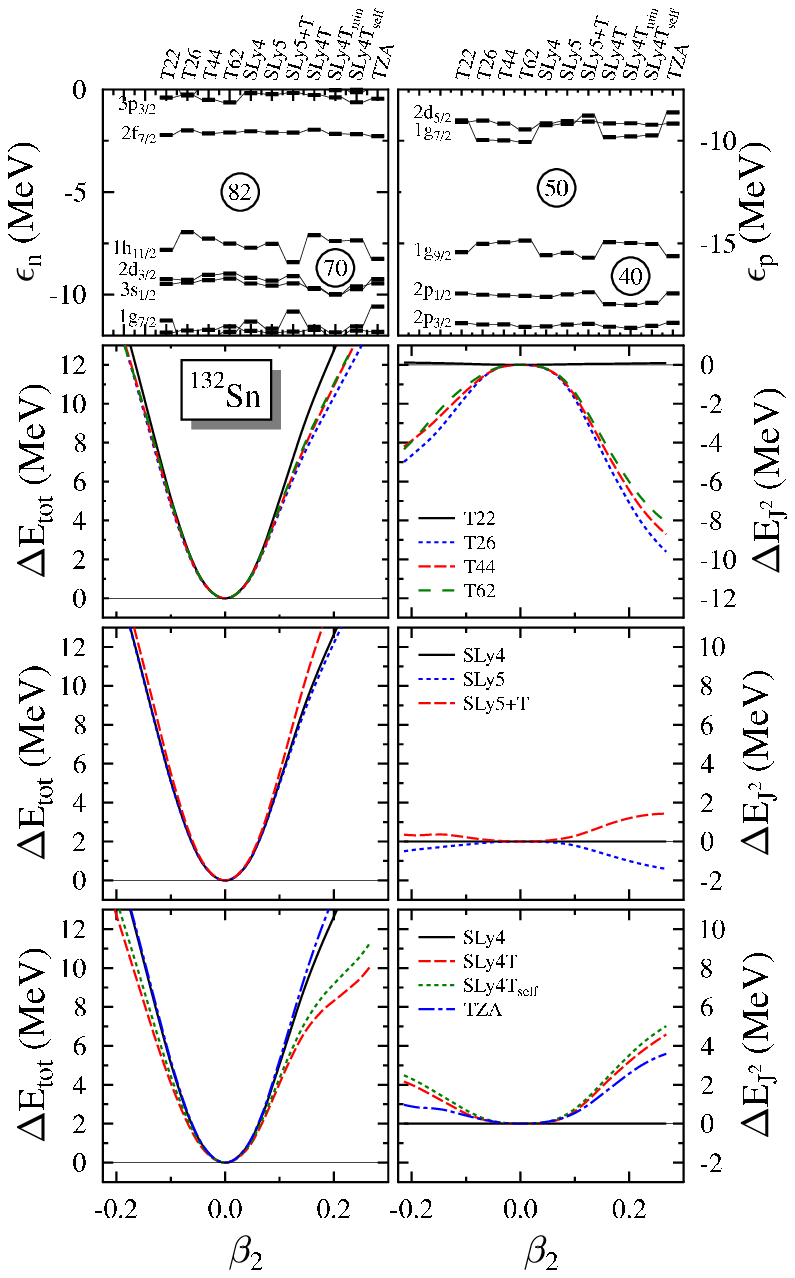}}
\caption{
\label{fig:sn132:surface}
(Color online)
Same caption as Fig.~\ref{fig:ni56:surface:comp1}, but for \nuc{132}{Sn}.
}
\end{figure}

\begin{figure}[t!]
\centerline{\includegraphics{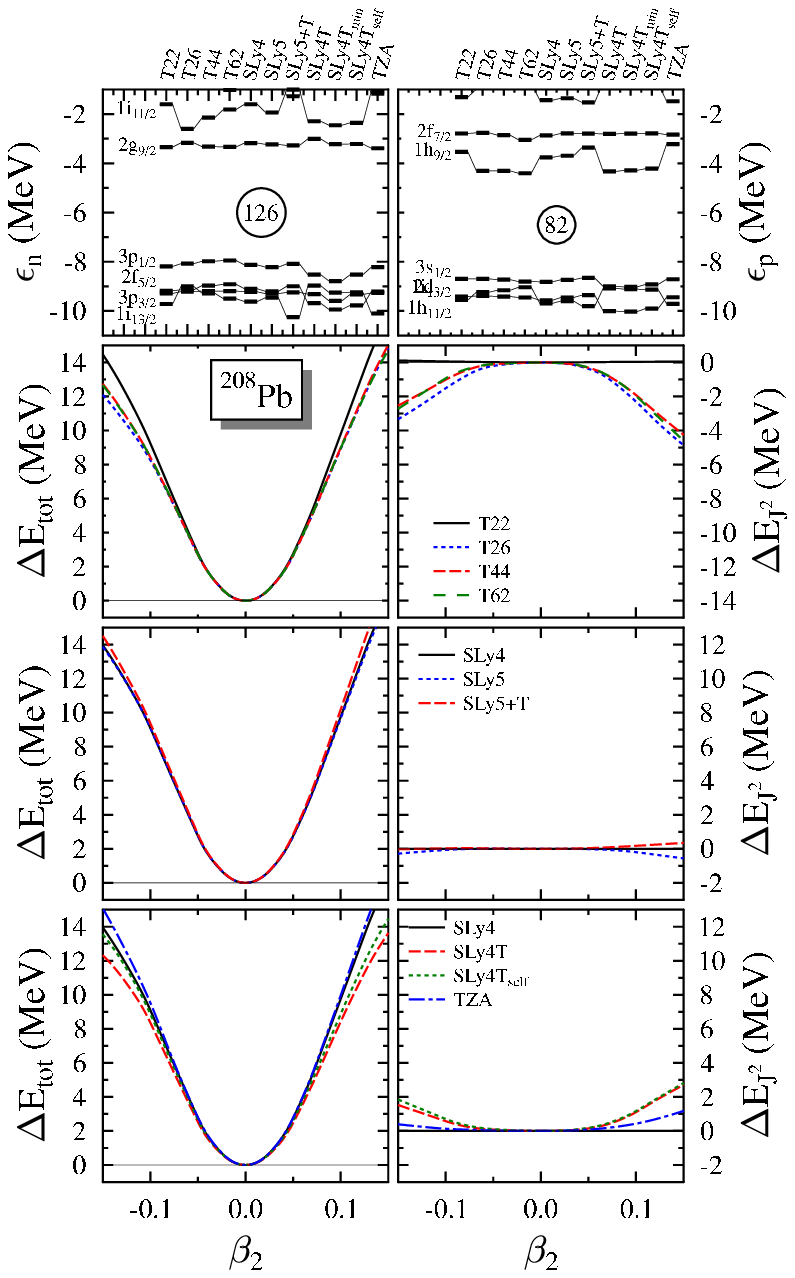}}
\caption{
\label{fig:pb208:surface}
(Color online)
Same caption as Fig.~\ref{fig:ni56:surface:comp1}, but for \nuc{208}{Pb}.
}
\end{figure}

%
%
\subsubsection{\nuc{132}{Sn} and \nuc{208}{Pb}}

The results obtained for the two heavy doubly-magic nuclei \nuc{132}{Sn}
and \nuc{208}{Pb} are presented in Figs.~\ref{fig:sn132:surface} and
\ref{fig:pb208:surface}. All neutron single-particle spectra
exhibit the usual problem of all mean-field interactions
that the $1h_{11/2}$ level in \nuc{132}{Sn} is not intruding the $gds$
shell~\cite{RMP}, as suggested by empirical data. The overall behavior
of the energy curves below 8~MeV is very similar for most interactions.
The stiffness of the deformation energy is marginally modified
by the tensor interaction and much less than one might have expected
from the variation of the single-particle spectra. The T$IJ$ interactions 
with $C^J_0$ coefficients different from zero give slightly softer
deformation energy curves than T22 or SLy4. However, the dependence
of the relative tensor energy on the value of the isovector coupling
constant $C^{J1}_1$ is very small for all deformations for
the same reason as for \nuc{78}{Ni}, in spite of the large asymmetry
of both nuclei. The reduction of the
spin-orbit strength for SLy4T and SLy4T$_{\text{self}}$ leads
to a prolate shoulder at about 10 MeV excitation energy.
As already found for lighter nuclei, the variation of the tensor energy
as a function of deformation can be large, up to 8~MeV in \nuc{132}{Sn}
and 4~MeV in \nuc{208}{Pb} for the rather small range of deformations covered
in Figs.~\ref{fig:sn132:surface} and~\ref{fig:pb208:surface}. This significant
variation is, to a large extent, absorbed by the rearrangement of the other
terms of the Skyrme functional, and it does not affect significantly the
total energy curves. The same mechanism that suppresses the isovector
tensor terms for \nuc{78}{Ni} is also at play in \nuc{132}{Sn}
and \nuc{208}{Pb}; hence, the variation of the tensor and total energy
with deformation of both nuclei is mainly correlated with the isoscalar
tensor coupling constant $C^J_0$.

\begin{figure}[t!]
\centerline{\includegraphics{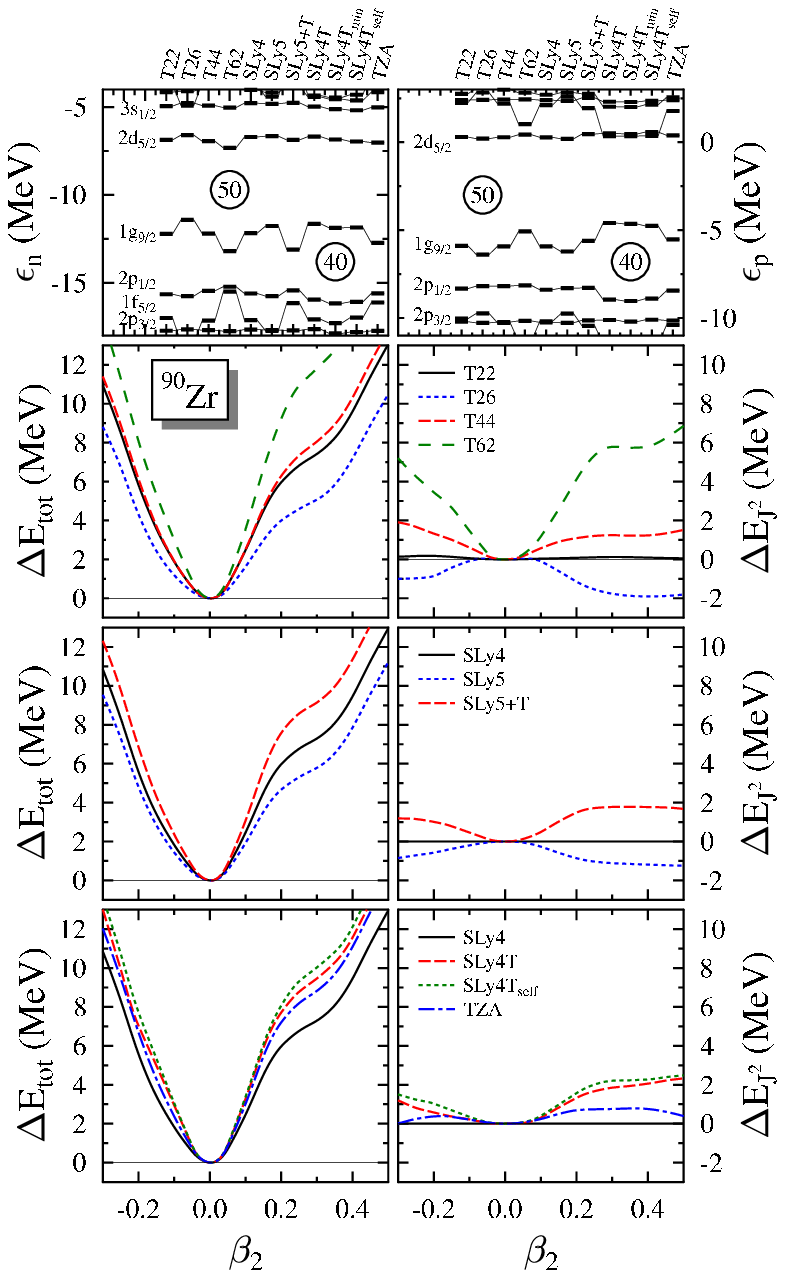}}
\caption{
\label{fig:zr90:surface}
(Color online)
Same caption as Fig.~\ref{fig:ni56:surface:comp1}, but for \nuc{90}{Zr}.
}
\end{figure}

%
%
\subsection{Selected Zr isotopes}

The Zr, $Z=40$, isotopic chain exhibits a rich spectroscopy, the
neutron-deficient and neutron-rich isotopes being very deformed
and the stable ones being spherical \cite{zrregion}. Self-consistent
mean-field methods experience large difficulties to reproduce these
very rapid variations of shapes in detail \cite{Ton81a,Ska93a,Rei99a}.
%
%
\subsubsection{\nuc{90}{Zr}}

The deformation energy and variation of the tensor contribution
with quadrupole deformation for \nuc{90}{Zr} are given in
Fig.~\ref{fig:zr90:surface}. There is a very close similarity
between the results obtained for \nuc{90}{Zr} and those for \nuc{48}{Ca},
see Fig.~\ref{fig:ca48:surface}.
In both cases, protons are spin-saturated at sphericity whereas neutrons 
are non-saturated as they fully occupy the lowest shell of a pair of 
spin-orbit partners.

\begin{figure}[t!]
\centerline{\includegraphics{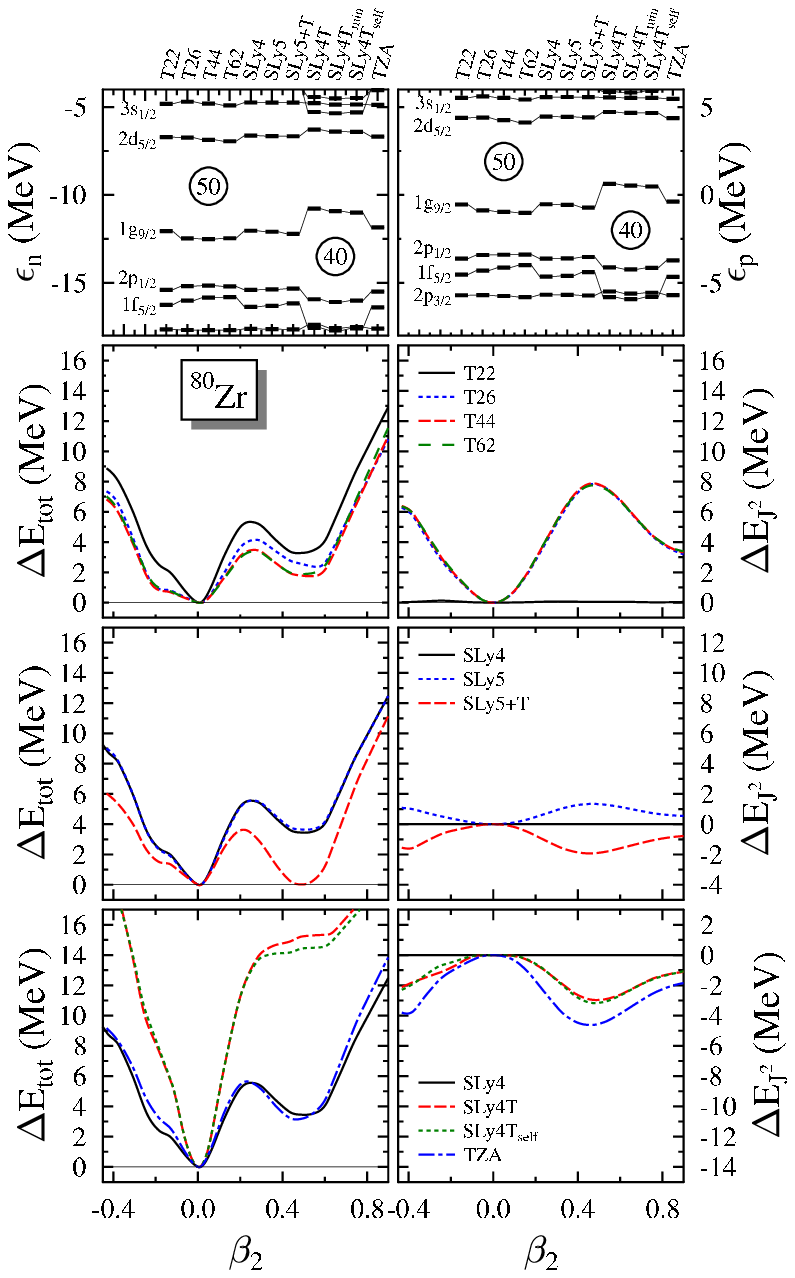}}
\caption{
\label{fig:zr80:surface}
(Color online)
Same caption as Fig.~\ref{fig:ni56:surface:comp1}, but for \nuc{80}{Zr}.
}
\end{figure}

%
%
\subsubsection{\nuc{80}{Zr}}

The situation is different for the $N = Z = 40$ isotope \nuc{80}{Zr}.
In spite of its double subshell closure, the sparse available spectroscopic
data suggest that \nuc{80}{Zr} has a large quadrupole deformation with a
$\beta_2$ value around 0.4. A rotational band built on the ground state
has been observed up to a spin of $10\hbar$~\cite{Lis87a,Lis90aE,Fis00a},
although it appears to be slightly distorted at low spin. The large 
deformation of states
in \nuc{80}{Zr} is also supported by the observation of strongly-coupled
rotational bands built on several Nilsson states in adjacent \nuc{79}{Y}
\cite{Pau98aE} and \nuc{81}{Zr} \cite{Mar04aE}. In the absence of information
on the transition matrix elements at the bottom of the band in \nuc{80}{Zr},
however, it is not ruled out that spherical and deformed configurations
might coexist in this nucleus and are strongly mixed in the ground state.

\begin{figure}[t!]
\centerline{\includegraphics{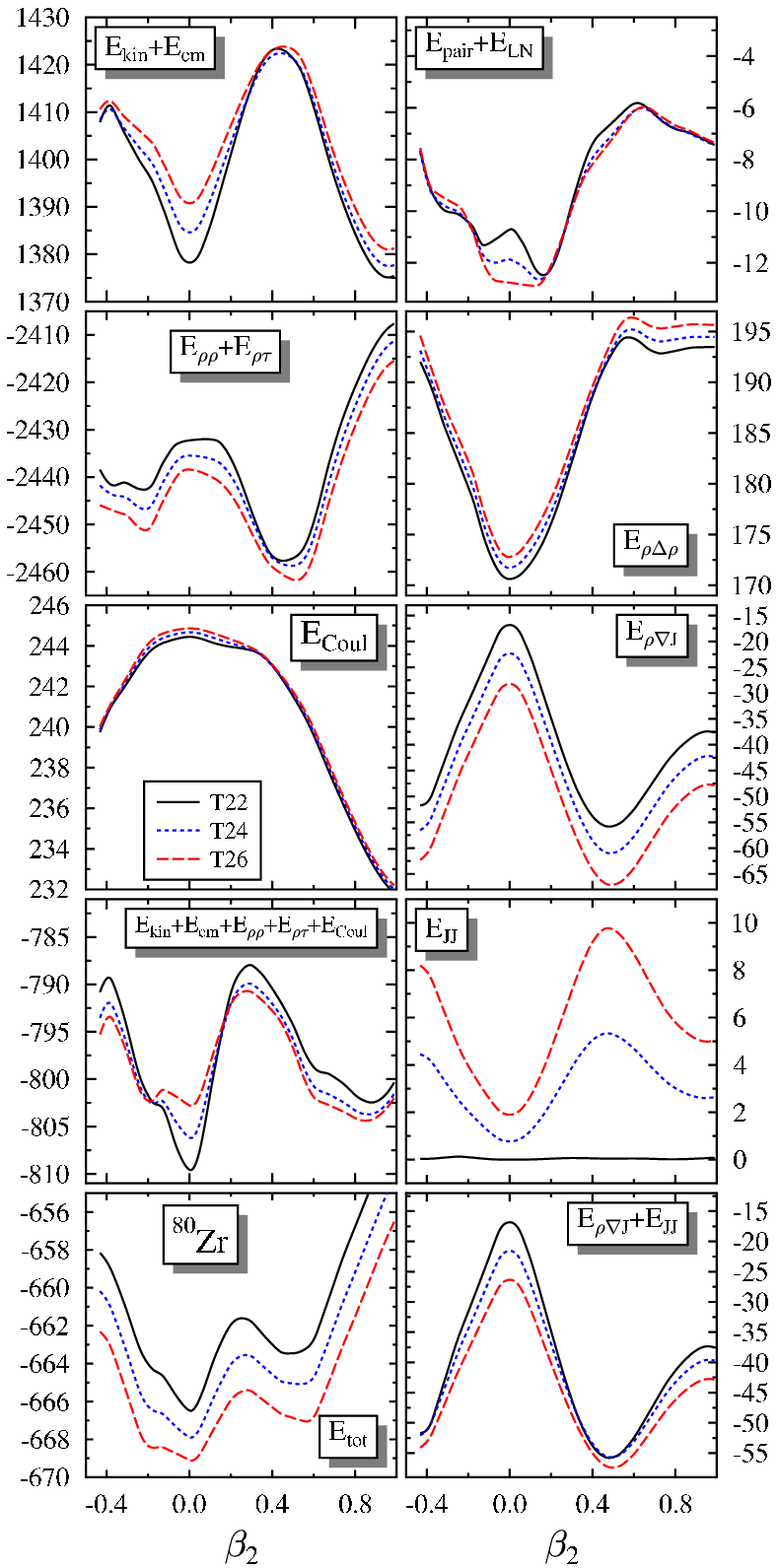}}
\caption{
\label{fig:zr80:total:decomposition}
(Color online)
Same caption as Fig.~\ref{fig:ni56:total:decomposition}, for \nuc{80}{Zr}.
}
\end{figure}

\begin{figure}[t!]
\centerline{\includegraphics{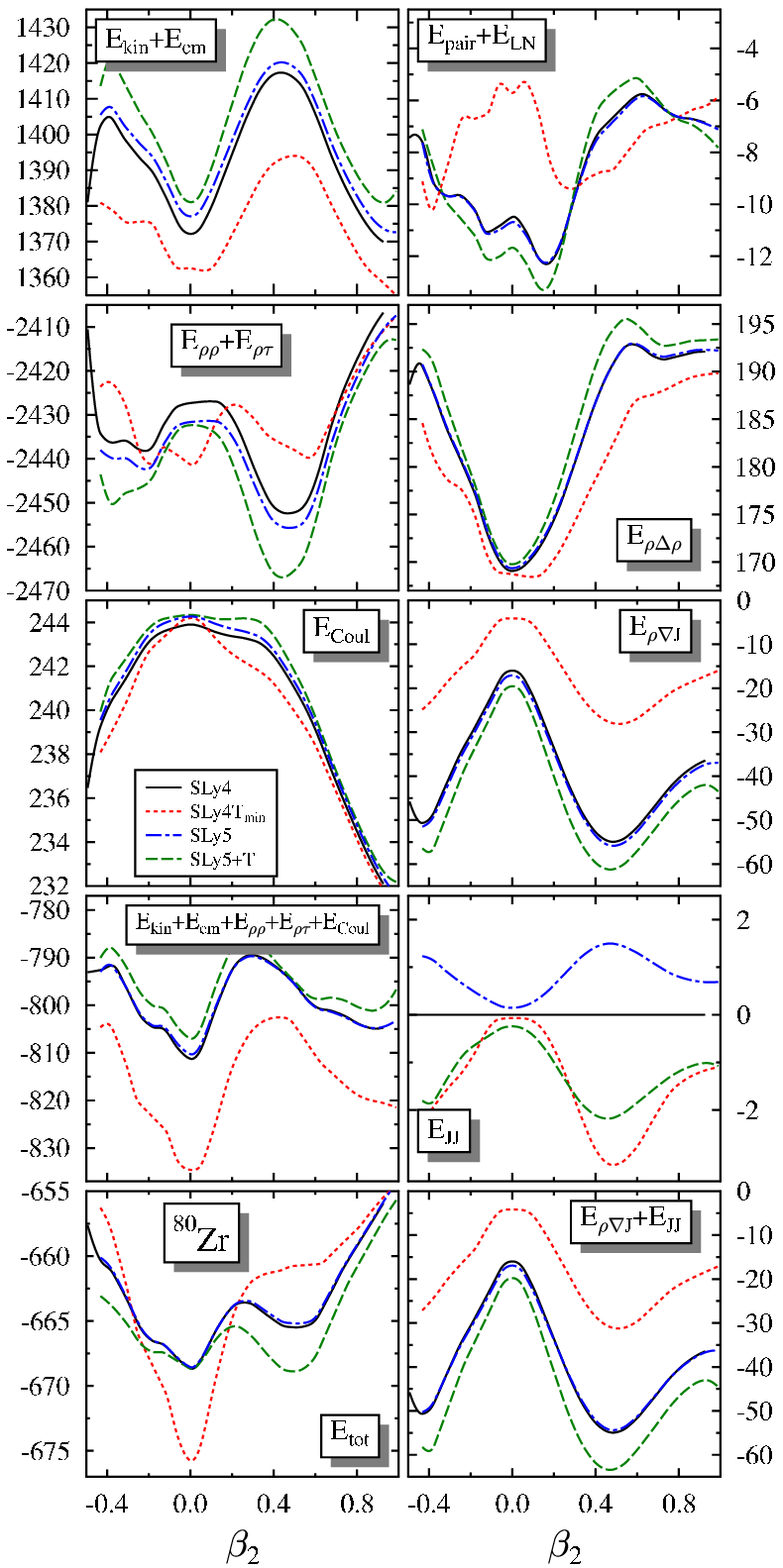}}
\caption{
\label{fig:zr80:total:decomposition2}
(Color online)
Same caption as Fig.~\ref{fig:ni56:total:decomposition2}, for \nuc{80}{Zr},
$Z = N = 40$.
}
\end{figure}

The deformation energy curves can be seen in Fig.~\ref{fig:zr80:surface}.
Protons and neutrons are spin-saturated at
sphericity. As a consequence, the predicted properties of this
nucleus present similarities with those of \nuc{40}{Ca}: the energy
of the tensor terms obtained using the T26, T44 and T62 interactions 
is very similar as only the isoscalar part of the tensor terms gives 
a sizable contribution. The total deformation energy, however, does 
exhibit a weak dependence on the value of $C^{J1}_1$. Also, 
comparing T22 and the other T$IJ$
interactions, the variation of the tensor terms with deformation is
in opposite direction to that of the total energy. Both results
illustrate the importance of the changes induced in all terms of
the functional by the fitting procedure.

The situation is different for interactions obtained by a perturbative
procedure. In this case, the addition of a tensor term to an existing
parameterization leads to more drastic changes. This is illustrated by
the comparison between the energy curves obtained with SLy5 and SLy5+T.
For the latter, the deformed minimum is pulled down and becomes degenerate
with the spherical configuration. The situation is opposite for SLy4 and
SLy4T. The large gap obtained with SLy4T for $Z=N=40$ has
a dramatic effect on the energy curve, which shows a sharp spherical minimum.
The large reduction of the spin-orbit strength is making this nucleus
doubly-magic and pushes the deformed minimum to a very high energy,
although the tensor interaction for SLy4T is more attractive for deformed
configurations. This effect is corrected for by TZA. However, none of
the parameterizations gives a deformed ground state, a deficiency shared
by many modern Skyrme interactions \cite{Rei99a}.

\begin{figure}[t!]
\centerline{\includegraphics{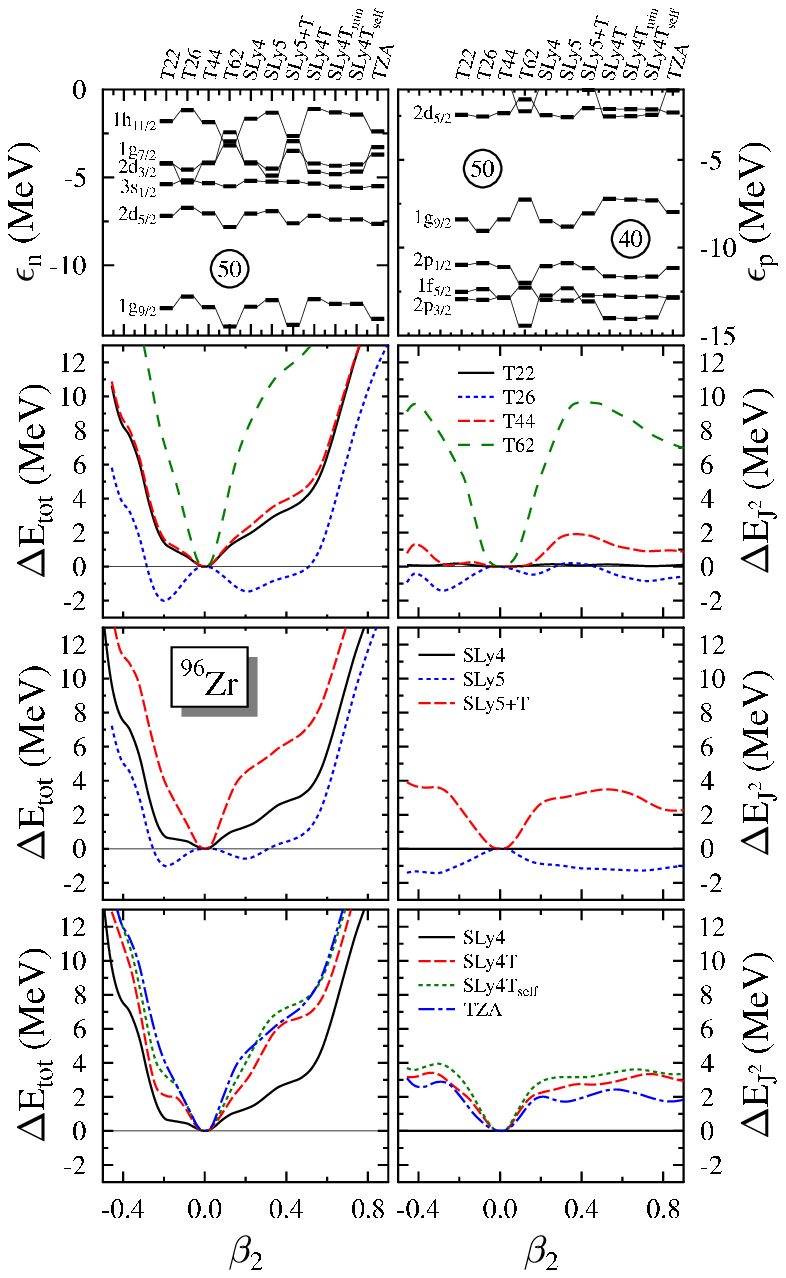}}
\caption{
\label{fig:zr96:surface}
(Color online)
Same caption as Fig.~\ref{fig:ni56:surface:comp1}, for \nuc{96}{Zr}.
}
\end{figure}

The decomposition of the energy into its central and spin-orbit+tensor
components is given in Fig.~\ref{fig:zr80:total:decomposition} for the
interactions T22, T24 and T26, and Fig.~\ref{fig:zr80:total:decomposition2}
for SLy4, SLy4T$_{\text{min}}$, SLy5 and SLy5+T.
They confirm the result found for \nuc{56}{Ni}, the topography
of the energy curves result from subtle cancellations between the bulk
contributions and the terms containing gradients. Again, the comparison of
different parameterizations indicates that the readjustment of the parameters
counteracts the self-consistency effects. The results obtained using
variational interactions are qualitatively very similar. On the contrary,
all components of the energy calculated with  perturbative
interactions are significantly different from those of the
original interaction.
A major qualitative difference with \nuc{56}{Ni} is that the bulk terms
give coexisting near-degenerate spherical and deformed minima in \nuc{80}{Zr},
and that the compensation between the gradient, spin-orbit and tensor terms
can tip the balance in one or the other direction.
A weak spin-orbit strength, such as for the SLy4T interaction, now has the 
effect of favoring the spherical minimum much too strongly, pushing the 
deformed minimum very high in energy.

\begin{figure}[t!]
\centerline{\includegraphics{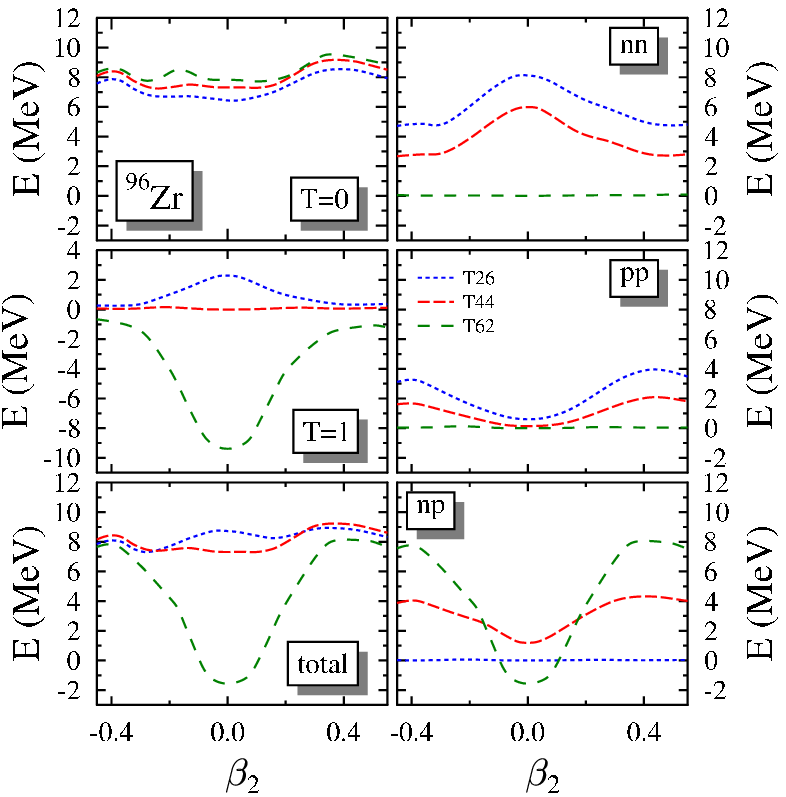}}
\caption{
\label{fig:zr96:tens:decomposition}
(Color online)
Same caption as Fig.~\ref{fig:ca48:tens:decomposition}, but for \nuc{96}{Zr}.
}
\end{figure}

%
%
\subsubsection{\nuc{96}{Zr}}

The \nuc{96}{Zr} isotope combines spherical sub-shell
closures at $N=56$ and $Z=40$. Its  low-energy spectrum exhibits several
unusual features. The systematics of masses in its immediate vicinity, 
a first $2^+$ level with a large excitation energy and one of 
the smallest $B(E2)$ values known
in heavy nuclei are all consistent with the expectation that \nuc{96}{Zr}
is a rigid spherical nucleus. Other observables
indicate the fragility of both shells. The $B(E3)$ value of the
$3^-_1 \to 0^+_1$ transition is among the strongest known
for a vibrational nucleus \cite{Mac90bE,Hor93aE}, the
charge radius is enhanced compared to the droplet-model trend \cite{Cam02aE},
thereby pointing to substantial ground-state correlations, and
the $g$ factors of the $2^+_1$ and $3^-_1$ hint at a
complex superposition of several neutron and proton excitations
across sub-shell closures~\cite{Kum03aE}. It also shares
with \nuc{90}{Zr} and \nuc{98}{Zr} the rare feature to have a low-lying
$0^+$ state as a first excited state~\cite{Mac88aE}. As for many
of the light doubly-magic nuclei studied above, this $0^+$
state is interpreted as a deformed state resulting from the simultaneous
2p-2h excitation of protons and neutrons across the respective
gaps~\cite{Hey88a}.

The single-particle spectra at spherical shape and the energy curves
of \nuc{96}{Zr} are presented in Fig~\ref{fig:zr96:surface}. As in \nuc{90}{Zr},
the neutrons are spin-unsaturated and the protons spin-saturated. However,
in this case, two levels contribute to the neutron spin-current density
at sphericity, $1g_{9/2^+}$ and $2d_{5/2^+}$; hence, this density is larger
than in \nuc{90}{Zr}. Thanks to that, the differences between the interactions
are amplified.
In particular, the contribution from the isovector tensor terms might
become very large as can be seen in Fig.~\ref{fig:zr96:tens:decomposition},
and drastically change the distance and even ordering of the single-particle
levels; see Fig.~\ref{fig:zr96:surface}. Compared to T22, T26 and T44,
the parameterization T62 gives much larger $N=56$ and $Z=40$ gaps, and
also pushes up the $2d_{3/2^+}$ and $1g_{7/2^+}$ neutron levels, both located 
above the Fermi energy. The SLy5+T interaction has the same tendency, but in
a less pronounced way. At least one of the tensor coupling constants
is negative for both interactions.

The deformed minima obtained with the T26 and SLy5 interactions 
are at variance with
data, as is the very stiff energy surface obtained with T62.

%
%
\subsubsection{\nuc{100}{Zr}}

\begin{figure}[t!]
\centerline{\includegraphics{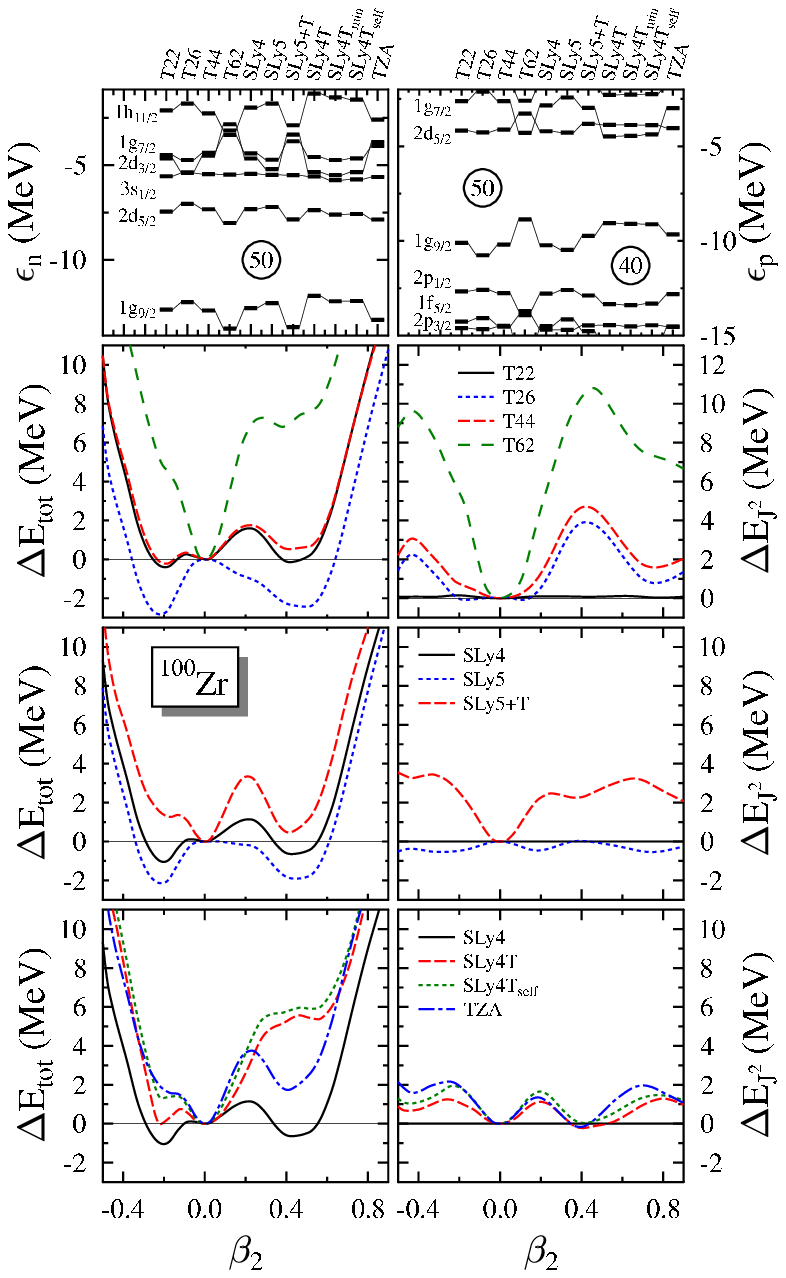}}
\caption{
\label{fig:zr100:surface}
(Color online)
Same caption as Fig.~\ref{fig:ni56:surface:comp1}, but for \nuc{100}{Zr},
$Z=40$, $N=60$.
}
\end{figure}

A large set of experimental data (charge radii~\cite{Cam02aE}, rotational
bands \cite{Hwa06bE} and $B(E2)$ values~\cite{Hwa06aE}) demonstrate that
\nuc{100}{Zr} is located in a region of deformed nuclei.
An excited band built on a $0^+$ state coexisting with the
ground-state band \cite{Hwa06bE}, and the large $E0$ transition strength
between the $0^+$ states \cite{Mac90aE} indicate the coexistence of shapes
with different deformations, the state with the largest deformation being
the ground state.

The single-particle spectra at spherical shape, the deformation energy
curve and the variation of the tensor energy are plotted against quadrupole deformation in
Fig.~\ref{fig:zr100:surface}. The overall behavior of the tensor energy
shows many similarities with \nuc{96}{Zr}. The results obtained with the T44
and T62 parameterizations indicate, however, a larger contribution from the 
isovector tensor
terms. The four additional neutrons shift the neutron  Fermi energy into
a region of large level density above the $N=56$ subshell closure.
The positions of the $2d_{3/2^+}$ and $2g_{7/2^+}$ are very much dependent
on the sign and size of the isovector coupling
constant $C^J_1$. For a positive value as in
T26, the $2d_{3/2^+}$ level is close to the Fermi level and is occupied in 
such a way that it partially cancels the contribution from the $2d_{5/2^+}$
orbital. The isovector tensor terms are in this case strongly reduced.
In contrast, for a negative $C^J_1$ coefficient as in T62, the $2d_{3/2^+}$ 
level is pushed up and crosses the $1h_{11/2^-}$ level,
increasing the neutron spin-current density. Results obtained with the
SLy5+T interaction, for which $C^J_1$ is also negative, are similar,
although less pronounced. For even larger negative values
of the tensor coupling  constants, this feedback mechanism will ultimately
generate an abnormal level ordering
for certain mid-shell nuclei, cf.\ the appendix~B of Article~I.
For SLy4T and SLy4T$_{\text{self}}$, this feedback mechanism is
suppressed by the reduced spin-orbit interaction, whereas for TZA, it is
present.

Most total deformation energy curves in Fig.~\ref{fig:zr100:surface}
exhibit spherical, prolate and oblate minima. The inclusion
of beyond mean-field correlations should favor the prolate minima and
create a $0^+$ excitation exhibiting some amount of configuration
mixing. Such results are consistent with experiment. The spherical minimum
is too much below the deformed one to expect that additional correlations
from the projection of $J=0$ states will make \nuc{100}{Zr} deformed in 
its ground state. For T62,  the deformation energy curve looks like that of a
doubly-magic nucleus. For SLy4T and SLy4T$_{\text{self}}$, it is the
reduced spin-orbit interaction that reinforces the proton $Z=40$ shell 
closure . The prolate minimum becomes a shoulder around 5~MeV, leading 
to the coexistence of spherical and oblate minima.

\begin{figure}[t!]
\centerline{\includegraphics{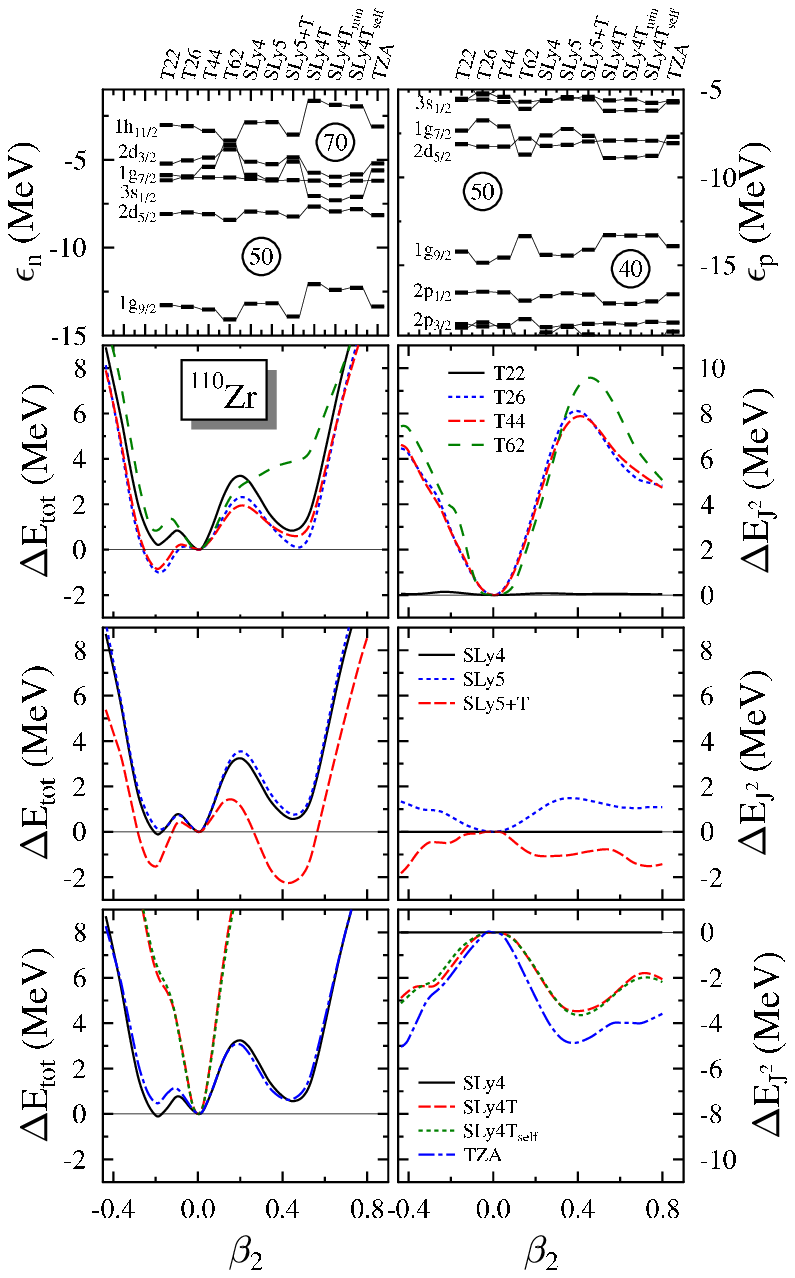}}
\caption{
\label{fig:zr110:surface}
(Color online)
Same caption as Fig.~\ref{fig:ni56:surface:comp1}, for \nuc{110}{Zr},
$Z=40$, $N=70$.
}
\end{figure}

%
\subsubsection{\nuc{110}{Zr}}

The only experimental information available about the very neutron-rich
\nuc{110}{Zr} is that it is a bound nucleus~\cite{Ber97aE}. It presents
the particularity to combine two spin-saturated oscillator shells, $Z=40$
and $N=70$. The corresponding gaps are still large for parameterizations
like SLy4T and SLy4T$_{\text{self}}$ with a reduced spin-orbit strength
and \nuc{110}{Zr} behaves like a doubly-magic nucleus. For all other
parameterizations, but T62, weak sub-shell closures remain at both
these nucleon numbers. The gaps are too small to enforce a rigid
spherical shape, but sufficient to prevent the existence of a clear-cut unique deformed minimum to describe the ground-state. Instead, all
interactions, but SLy4T, SLy4T$_{\text{self}}$ and T62, predict a complicated
pattern of three coexisting spherical, prolate and oblate structures. For
T62, there is no prolate minimum, and the spherical configuration is favored
because of the semi-magic character of this nucleus with a large $Z=40$
shell closure.
For SLy4T and SLy4T$_{\text{self}}$, there is a single, very sharp spherical minimum typical of a doubly-magic character.

%
%
\subsection{Heavy semi-magic nuclei}

Let us conclude our survey with two selected heavy semi-magic nuclei.

%
\subsubsection{\nuc{120}{Sn}}

\begin{figure}[t!]
\centerline{\includegraphics{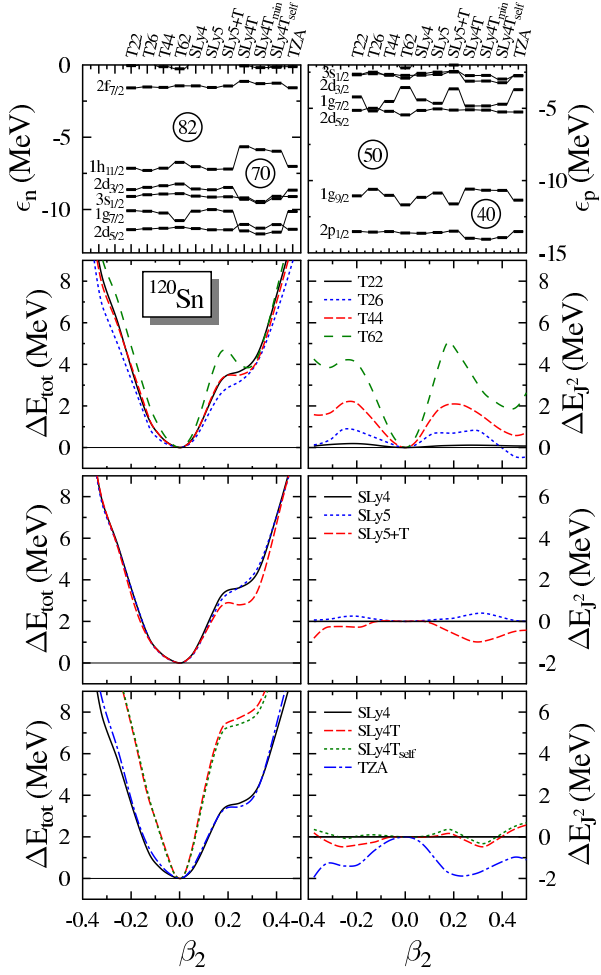}}
\caption{
\label{fig:sn120:surface}
(Color online)
Same caption as Fig.~\ref{fig:ni56:surface:comp1}, for \nuc{120}{Sn},
$Z=50$, $N=70$. All panels share the same energy scale.
}
\end{figure}

\begin{figure}[t!]
\centerline{\includegraphics{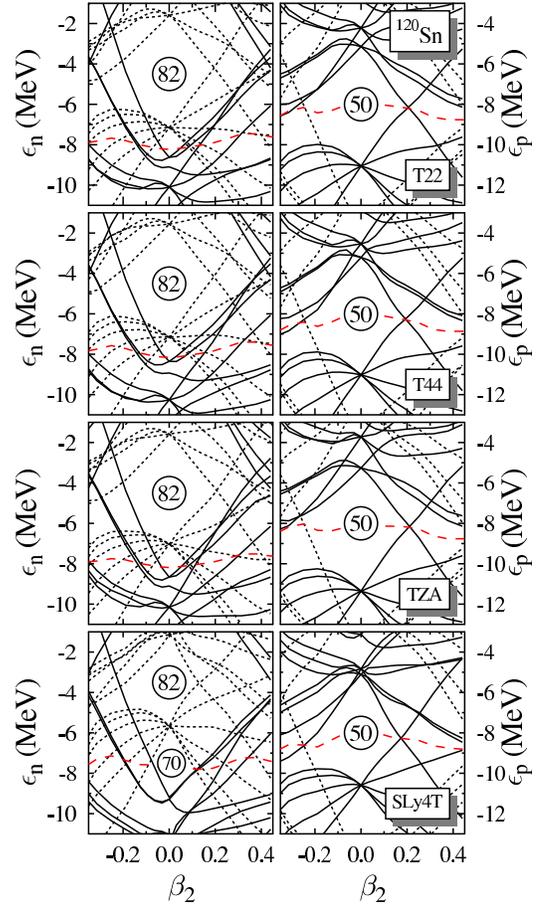}}
\caption{
\label{fig:sn120:nilsson}
(Color online)
Proton and neutron Nilsson diagrams for \nuc{120}{Sn} obtained with the parameterizations
as indicated. Solid lines denote levels of positive parity, dotted
lines levels of negative parity, and the red dashed lines denote the
Fermi energy.
}
\end{figure}

The stable semi-magic \nuc{120}{Sn} is the lightest of heavy tin
isotopes for which no coexisting deformed rotational band at low
excitation energy has been observed \cite{Jul95aR}.
The neutron number $N=70$ of \nuc{120}{Sn} corresponds to a
magic number for neutrons in a pure harmonic oscillator picture.
This simple picture is destroyed by the spin-orbit interaction
which pushes the $1h_{11/2^-}$ across the $N=70$ gap, creating a
shell closure at $N=82$. In fact, data suggest that this
oscillator shell does not survive even as a subshell
closure, as the empirical $11/2^-$ intruder level is
below the $3/2^+$ level and degenerate with the $1/2^+$ state in
\nuc{132}{Sn}. As already mentioned above, it is a well-known problem
of virtually all energy functionals that the $1h_{11/2^-}$ intruder
level is predicted to lie slightly above the $gds$ shell~\cite{RMP}.
This deficiency was related in Fig.~17 of Article~I to a too high
position of the centroid of the $1h$ levels.

The energy surfaces obtained with the T$IJ$ interactions are presented 
on the left-hand side of Fig.~\ref{fig:sn120:surface}, the Nilsson diagrams
for four selected parameterizations in Fig.~\ref{fig:sn120:nilsson}.
The neutron contribution to the tensor energy is small at sphericity,
as the neutrons are predicted to be spin saturated, see
Fig.~\ref{fig:sn120:nilsson}, at variance with experiment.
As soon as deformation sets in, the tensor energy increases for
the four T$IJ$ parameterizations. However, the total energy curves
are much closer than one would expect from the difference
between the tensor energies. The most significant difference is obtained
for deformations between the spherical minimum and the prolate shoulder.

The energy curves calculated with the SLy4, SLy5 and SLy5+T 
interactions are nearly identical, except for a small lowering of the 
prolate shoulder for the latter.

The situation is quite different for SLy4T. As a consequence of its
weak spin-orbit strength, the neutron intruder level
is halfway in the gap between
the major shells. The energy gap at $N=70$ remains very large
and significantly reduces the neutron level density around the
Fermi energy for deformations up to $\beta_2$ values around
0.1. The tensor energy decreases with deformation,
but not sufficiently to compensate the effect of the decrease of the
spin-orbit strength. The net effect on the energy curve is that it is
much stiffer than with the original SLy4 parameters, artificially
making \nuc{120}{Sn} a doubly-magic nucleus similar to \nuc{132}{Sn}.

%
%
\subsubsection{\nuc{186}{Pb}}

\begin{figure}[t!]
\centerline{\includegraphics{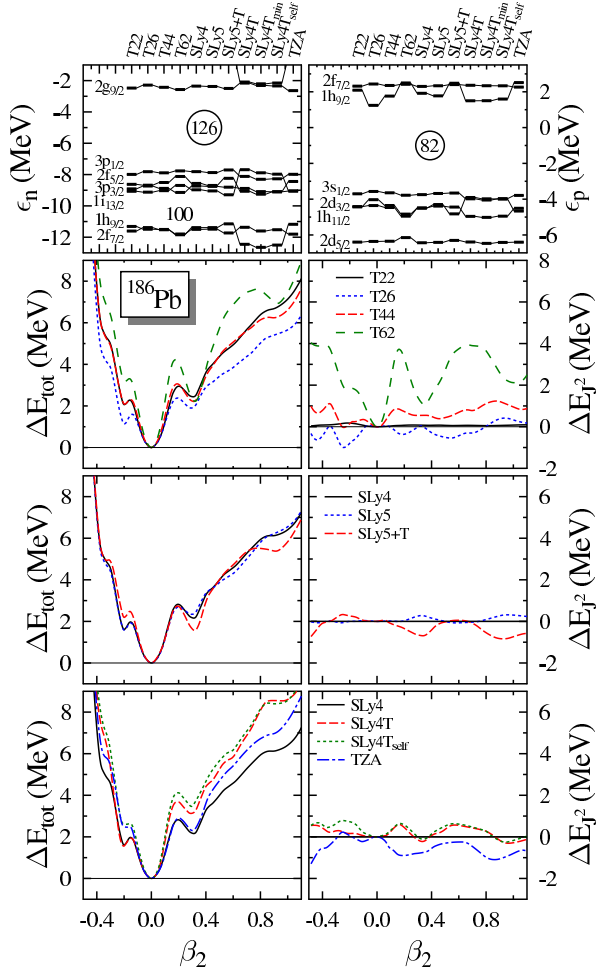}}
\caption{
\label{fig:pb186:surface}
(Color online)
Same caption as Fig.~\ref{fig:ca40:surface}, but for \nuc{186}{Pb},
$Z=82$, $N=104$.
}
\end{figure}

The heavy, neutron-deficient $N=104$, $Z=82$ Pb isotope, \nuc{186}{Pb},
exhibits a triple shape coexistence of spherical, prolate and oblate
shapes, with the unique feature that its two lowest excited levels
are $0^+$ states~\cite{Andreyev00,Andreyev01}. The
deformation energy curves obtained with  all interactions tested here
are plotted in Fig.~\ref{fig:pb186:surface}. They are compatible with
the experimental data and present a spherical minimum,
and excited oblate, prolate, and often also superdeformed minima,
in most cases all separated by small barriers.

The tensor energy and the impact of the tensor terms on the total energy
are similar to those found for \nuc{120}{Sn}. The differences between 
the T$IJ$ interactions are the largest between the minima or shoulders;
SLy5+T slightly moves the excitation energies of excited minima compared to
SLy5, and the deformation energy curves from SLy4T and SLy4T$_{\text{self}}$
are stiffer than the others,
at least for prolate deformations. The reduced spin-orbit strength for
SLy4T and SLy4T$_{\text{self}}$ pulls the neutron $1i_{13/2^+}$ intruder
back towards the $N=126$ gap. This is inconsistent with the existence of a
very low-lying isomeric $13/2^+$ states located at a few tens of keV 
excitation energy in surrounding odd-$A$ Pb isotopes. $\alpha$-decay
hindrance factors suggest indeed that this state is well described
by a neutron in the $1i_{13/2^+}$ level coupled to a spherical
core~\cite{And99aE}.

Nuclei in this mass region are less affected by the tensor terms, as
hinted already in Article~I by the analysis of the spin-current
density at spherical shape in the Pb isotopic chain.
Still, the tensor terms modify the balance between the excitation energy of
the coexisting minima. As the relative position of the minima is sensitive
to all terms of the EDF, this quantity cannot be used to
safely validate the tensor coupling constants.

%
%

\section{Summary and Conclusions}

We have studied the impact of tensor terms in the
Skyrme energy density functional on deformation properties of magic and 
semi-magic nuclei. This work is a continuation of a previous study limited 
to spherical symmetry, as published in Article~I~\cite{Les07a}.
The study has been focussed on a representative sample of
parameterizations introduced in Article~I, which covers a
wide range of values for the isoscalar and isovector tensor coupling constants
and allow to disentangle their respective role. These parameterizations 
are adjusted with a fit protocol very similar to that of the successful 
SLy$x$ parameterizations \cite{Cha97a,Cha98a}. We also considered two other 
recent families of energy functionals also based on the SLy$x$ ones, but 
constructed following very different strategies. For the parameterization 
SLy5+T \cite{Col07a} a tensor force was perturbatively added to SLy5 
without any readjustment of the other parameters. For SLy4T \cite{Zal08a},
the tensor and spin-orbit coupling constants were fixed without any 
readjustment of the other parameters of SLy4. The related parameterizations
SLy4T$_{\text{min}}$ \cite{Zal08a} and SLy4T$_{\text{self}}$ and TZA 
introduced here allow to disentangle the origin of the different results 
obtained with SLy4T and the T$IJ$s as being due to the perturbative fit, 
the change in the spin-orbit strength, or the choice of tensor coupling
 constants.

A first result that we have obtained concerns the order of magnitude of the
different components of the tensor term. In spherical coordinates, it can 
be decomposed in vector and pseudotensor contributions. For all studied 
parameterizations for which the coefficients of both terms have the same 
order of magnitude, the pseudotensor contribution is at least one order 
of magnitude lower than the vector one. This justifies the common practice 
of neglecting the pseudovector contribution to the energy.

The  shell effects induced by the tensor interaction fluctuate as a function 
of the filling of single-particle orbits. This effect has motivated the 
introduction of the tensor force to explain the evolution of the shell 
structure of spherical nuclei along isotopic lines. Inevitably, it leads 
also to a pattern for the size and deformation dependence of the 
contribution of the tensor terms to the total energy which depends on 
the fillings of orbitals:
\begin{itemize}
\item[(i)]
For doubly spin-saturated nuclei at sphericity, such as \nuc{40}{Ca} and
\nuc{80}{Zr}, the tensor energy is close to zero at spherical shape, and
increases in absolute value with deformation.
\item[(ii)]
For doubly spin-unsaturated doubly-magic nuclei such as \nuc{56}{Ni},
\nuc{78}{Ni}, \nuc{100}{Sn}, \nuc{132}{Sn} and \nuc{208}{Pb},
the absolute value of the tensor energy is the largest at sphericity,
and decreases with deformation.
\item[(iii)]
For doubly spin-unsaturated doubly-magic $N=Z$ nuclei like \nuc{56}{Ni}
and \nuc{100}{Sn}, the tensor energy is obviously dominated by
the isoscalar part of the tensor interaction. The same conclusion holds,
however, also for the $N \neq Z$ nuclei \nuc{78}{Ni}, \nuc{132}{Sn}
and \nuc{208}{Pb}, in spite of their large asymmetry $N-Z$.
The reason for that is that the proton and neutron spin-currents 
densities are very
similar in size, sign and spatial distribution in these nuclei; hence,
they nearly cancel each others' contribution to the isovector spin-current
at all deformations.
\item[(iv)]
The isovector tensor contribution
to the energy
plays a significant role
only for doubly-magic nuclei that combine a spin-saturated
configuration for one nucleon species with a spin-unsaturated
configuration for the other, such as in \nuc{48}{Ca}, \nuc{68}{Ni}, and
\nuc{90}{Zr}.
\item[(v)]
The behavior of nuclei without large shell or sub-shell closures for
at least one nucleon species does not follow simple rules.
These nuclei are most sensitive
to the values of the tensor coupling constants, at least within the
sample of nuclei studied here. In nuclei with a large density of
single-particle levels around the Fermi surface, there are highly
nonlinear feedback effects at play. For large absolute values of
their coupling constants, the tensor terms reduce or amplify themselves
through the reordering of levels around the Fermi energy, as exemplified
by \nuc{96}{Zr} and \nuc{100}{Zr}.
\end{itemize}
Self-consistency is implemented at two different levels in the method
that we have used: in the fitting procedure of the interaction and in the
solution of the mean-field equations.
\begin{itemize}
\item[(i)]
The perturbative addition of only a tensor term, like for SLy5+T,
to an existing parameterization  will modify all contributions to the 
mean fields and the energy.
\item[(ii)]
The self-consistency of the mean-field
induces a rearrangement of the single-particle wave functions, and
consequently of all densities affecting at the end all observables. 
This effect is exemplified by the comparison between the results 
obtained with SLy5 and SLy5+T, which share all coupling constants except 
those of the tensor terms.
\item[(iii)]
Using a protocol mainly based on infinite nuclear matter properties,
binding energies and charge radii, as the Saclay-Lyon 
protocol~\cite{Cha97a,Cha98a}, the changes in the coupling constants 
due to the self-consistency of fits tend to counteract the self-consistency 
in the mean field. This is exemplified by the comparison between the results 
obtained with SLy4, SLy5 and SLy5+T. For most
nuclei studied here and for most quantities not directly affected by the
tensor  terms, the differences between the predictions of the first two are
on a much smaller scale than the differences between the latter
two. A perturbative modification of a well-adjusted parameterization might
spoil its predictive power in unexpected ways. Our results confirm the
suspicion of the authors of the perturbatively constructed SLy5+T
\cite{Col07a}, who indeed intended their interaction as a tool for
explorative studies only, and state that "an ambitious refitting program 
[\ldots] should be [\ldots] undertaken" for more detailed studies.
\item[(iv)]
Self-consistency of the fits and/or the calculations has the consequence
that the total deformation energy obtained with different interactions
varies in most cases on a much smaller scale than the tensor contributions.
In some cases such as \nuc{80}{Zr}, they even might go into opposite 
directions.
\item[(v)]
The tensor and spin-orbit contributions to the total energy and
to the spin-orbit fields  are tightly interwoven.
Constraining both too tightly in a small region of the nuclear chart
might be misleading when aiming at a universal functional.
This is exemplified by SLy4T with its spin-orbit and tensor coupling
constants fitted very carefully to suitably chosen spin-orbit splittings
in \nuc{40}{Ca}, \nuc{48}{Ca} and \nuc{56}{Ni}. The failure of SLy4T
to extrapolate well clearly points to missing physics, either in the
form of missing terms in the functional or missing correlations.
\item[(vi)]
The strong reduction of the spin-orbit strength for SLy4T improves
the description of spin-orbit splittings in light nuclei, but
amplifies the problems from the wrong positioning of centroids.
Also, the spin-orbit splittings in heavy nuclei are much too small.
The strong reduction of the spin-orbit strength to about $2/3$
its original value is specific to the SLy4-based interaction
constructed in~\cite{Zal08a}. For their SkP and SkO based fits, the
reduction is much more moderate.
\end{itemize}
The size and deformation dependence of the tensor
energy is correlated with the impact of the tensor terms on single-particle
spectra. In Article~I, we analyzed how the tensor terms affect the position
and relative distance of single-particle energies for spherical shapes.
Concerning the dependence of Nilsson diagrams on the tensor force, the 
following can be stated:
\begin{itemize}
\item[(i)]
Tensor terms modify the slope of the levels in the Nilsson diagram.
For the magic nuclei studied here this happens in particular around
sphericity, where the tensor contribution to the spin-orbit field
$W_{q,\mu \nu}$ often changes rapidly with deformation.
\item[(ii)]
When comparing interactions with different values of the tensor
coupling constants that are otherwise completely refitted,
the  change in slope compensates at large deformations to a large
extent the differences between single-particle spectra found at spherical
shape. For those interactions, the deformed
single-particle spectra around the Fermi energy are often nearly identical
in spite of the different tensor interactions. In such fit protocol, the
coupling constants of the tensor terms control the balance between
spherical and deformed shell gaps.
\item[(iii)]
In perturbative fits, in particular those where more than one term is
rescaled, deformed shell structure is affected as well.
\end{itemize}
It can be expected that these finding are to a large extent independent
of remaining deficiencies of the central and spin-orbit interactions,
and will be of great value for the construction of future, improved
energy functionals. We will address the question of how the surface and
surface symmetry energy coefficients of the interactions change as a function
of the coupling constants of the tensor terms, and how this correlates
with energy at large deformation in future work. A study of
the so-called "time-odd terms" in the energy functional that originate
from a tensor force is underway as well. A point of special interest
will be the analysis of potential finite-size instabilities using the
technique of Ref.~\cite{Les06a,Dav09a}.

%
%
\acknowledgments

This work was supported by the "Interuniversity Attraction Pole"
(IUAP) of the Belgian Scientific Policy Office under project P6/23; 
by the U.S.\ National Science Foundation under Grant No\. PHY-0456903, 
and by the U.S.\ Department of Energy under Contract Nos.\ 
DE-FG02-96ER40963, DE-FG02-07ER41529 (University of 
Tennessee) and DE-AC05-00OR22725 with UT-Battelle, LLC (Oak Ridge 
National Laboratory).

%
%
\bibliography{bender_tens_def_v1}

\end{document}